\documentclass{emulateapj}

\newcommand{\hvc}{HV~Tau~C}
\newcommand{\hv}{HV~Tau}
\newcommand{\hvab}{HV~Tau~AB}
\newcommand{\hva}{HV~Tau~A}
\newcommand{\hvb}{HV~Tau~B}
\newcommand{\mic}{$\mu$m}

\shorttitle{Observations and modeling of HV\,Tau\,C}
\shortauthors{Duch\^ene et al.}

\begin{document}

\title{Panchromatic observations and modeling of the HV\,Tau\,C
  edge-on disk\footnote{Data presented in this study were obtained
    during the course of ESO program 70.C-0565 and IRAM program
    O048.}}

\author{G. Duch\^ene\altaffilmark{1}} \affil{Astronomy Department,
  University of California, Berkeley, CA 94720-3411, USA}
\email{gduchene@astro.berkeley.edu}

\and

\author{C. McCabe}\affil{IPAC, MS 220-6, California Institute of
  Technology, Pasadena, CA 91125, USA}

\and

\author{C. Pinte\altaffilmark{1}}\affil{School of Physics, University
  of Exeter, Stocker Road, Exeter EX4 4QL, UK}

\and

\author{K. R. Stapelfeldt}\affil{JPL, MS 183-900, California Institute
  of Technology, Pasadena, CA 91109-8099, USA}

\and

\author{F. M\'enard} \author{G. Duvert} \affil{Universit\'e Joseph
  Fourier - Grenoble 1 / CNRS, laboratoire d'Astrophysique de Grenoble
  (LAOG) UMR 5571, BP 53, 38041 Grenoble Cedex 09, France}

\and

\author{A. M. Ghez}\affil{Division of Astronomy and Astrophysics,
  UCLA, Box 951547, Los Angeles, CA 90095-1562, USA}

\and

\author{H. L. Maness} \affil{Astronomy Department, University of
  California, Berkeley, CA 94720-3411, USA}

\and

\author{H. Bouy}\affil{Instituto de Astrofisica de Canarias, C/ Via
  Lactea s/n, 38200 - La Laguna, Tenerife, Spain}

\and

\author{D. Barrado y Navascu\'es} \author{M. Morales-Calder\'on}
\affil{Laboratorio de Astrofisica Estelar y Exoplanetas (LAEX-CAB,
  INTA-CSIC) P.O.Box 78, 28691 Villanueva de la Canada (Madrid), Spain}

\and

\author{S. Wolf} \affil{Christian-Albrechts-Universit\"at zu Kiel,
  Institut f\"ur Theoretische Physik und Astrophysik, Leibnizstr. 15,
  24098 Kiel, Germany}

\and

\author{D. L. Padgett} \author{T. Y. Brooke}
\author{A. Noriega-Crespo} \affil{SSC, MS 220-6, California Institute
  of Technology, Pasadena, CA 91125, USA}

\altaffiltext{1}{Universit\'e Joseph Fourier - Grenoble 1 / CNRS,
  laboratoire d'Astrophysique de Grenoble (LAOG) UMR 5571, BP 53,
  38041 Grenoble Cedex 09, France}

\begin{abstract}
  We present new high spatial resolution ($\lesssim0$\farcs1)
  1--5\mic\ adaptive optics images, interferometric 1.3\,mm continuum
  and $^{12}$CO 2--1 maps, and 350\mic, 2.8 and 3.3\,mm fluxes
  measurements of the {\hv} system. Our adaptive optics images
  unambiguously demonstrate that {\hvab}--C is a common proper motion
  pair. They further reveal an unusually slow orbital motion within
  the tight {\hvab} pair that suggests a highly eccentric orbit and/or
  a large deprojected physical separation. Scattered light images of
  the {\hvc} edge-on protoplanetary disk suggest that the anisotropy
  of the dust scattering phase function is almost independent of
  wavelength from 0.8 to 5\mic, whereas the dust opacity decreases
  significantly over the same range. The images further reveal a
  marked lateral asymmetry in the disk that does not vary over a
  timescale of 2 years. We further detect a radial velocity gradient
  in the disk in our $^{12}$CO map that lies along the same position
  angle as the elongation of the continuum emission, which is
  consistent with Keplerian rotation around an 0.5--1$M_\odot$ central
  star, suggesting that it could be the most massive component in the
  triple system. To obtain a global representation of the {\hvc} disk,
  we search for a model that self-consistently reproduces observations
  of the disk from the visible regime up to millimeter wavelengths. We
  use a powerful radiative transfer model to compute synthetic disk
  observations and use a Bayesian inference method to extract
  constraints on the disk properties. Each individual image, as well
  as the spectral energy distribution, of {\hvc} can be well
  reproduced by our models with fully mixed dust provided grain growth
  has already produced larger-than-interstellar dust grains. However,
  no single model can satisfactorily simultaneously account for all
  observations. We suggest that future attempts to model this source
  include more complex dust properties and possibly vertical
  stratification. While both grain growth and stratification have
  already been suggested in many disks, only a panchromatic analysis,
  such as presented here, can provide a complete picture of the
  structure of a disk, a necessary step towards quantitatively testing
  the predictions of numerical models of disk evolution.
\end{abstract}


\section{Introduction}

Circumstellar disks are an ubiquitous outcome of the stellar formation
process and they are believed to be the birth place of planetary
systems. The growth of dust particles towards planetesimal sizes along
with their vertical settling due to gas drag are processes that are
believed to be the first steps towards planet
formation. Hydrodynamical models have shown that these processes can
be efficient early in the disk evolution
\citep[e.g.,][]{weidenschilling97, dullemond04, barriere05}. To test
these models, it is necessary to obtain an observation-based
description of the structure and dust content of protoplanetary disks
as a function of the age of the system and other relevant parameters
(e.g., stellar mass).

The dust component of protoplanetary disks has long been studied via
its thermal emission from near-infrared to millimeter wavelengths
which is frequently associated with low-mass pre-main sequence
T\,Tauri stars \citep{kenyon87, bertout88, strom89, beckwith90}. Both
grain growth and dust settling can alter the overall shape of the
spectral energy distribution (SED) of a T\,Tauri star
\citep{dalessio01, dullemond04, dalessio06}. Indeed, several studies
that analyzed (elements of) the SED of young stars have concluded that
both grain growth and settling is occuring in protoplanetary disks
\citep[e.g.,][and references therein]{beckwith91, mannings94,
  furlan06, kessler06, rodmann06, natta07}. Unfortunately, such
studies suffer from the absence of spatial information inherent to
photometric measurements and the high optical thickness of disks in
the near- to mid-infrared regime. As a result, comparing an object's
SED to radiative transfer models leaves many ambiguities
\citep{chiang01}. For instance, the inferred total dust mass and the
maximum size of the dust grains are inversely correlated because of
the dependency of dust opacity on grain size. In addition, most of
these studies, which focus on a single type of observations (e.g.,
millimeter fluxes, silicate emission feature), only probe a limited
region of the disk and a small fraction of the entire grain size
distribution.

To solve for the ambiguities inherent to SED studies, it is critical
to obtain spatially-resolved observations. Such observations include
thermal emission mapping with (sub)millimeter interferometers
\citep[e.g.,][]{keene90, simon92, lay94, dutrey96, andrews07} and
scattered light imaging with optical and near-infrared high-resolution
instruments \citep[e.g.,][]{burrows96, roddier96,
  stapelfeldt98}. Disks are generally optically thin at long
wavevengths, so the former type of observations can probe the entire
disk structure. Furthermore, they are very sensitive to the presence
of millimeter-sized particles. On the other hand, scattered light
images, which only probe dust grains at the disk surface, are very
sensitive to the size distribution of micronic grains especially when
images at multiple wavelengths are analyzed simultaneously
\citep[][and references therein]{watson07ppv}. Both types of
observations have already yielded important pieces of evidence
supporting both grain growth and dust settling in disks
\citep[e.g.,][]{duchene03, duchene04, watson04}.

While grain growth and settling appear to occur in protoplanetary
disks, detailed {\it quantitative} tests of hydrodynamical models can
only be achieved with a detailed view of the entire structure of a
disk. This can only be obtained via a multi-technique, panchromatic
approach. Unfortunately, observational and computational limitations
have so far limited the number of objects for which such an analysis
could be conducted to a handful. The most notable examples are the
studies of the ``Butterfly Star'' \citep{wolf03}, IM\,Lup
\citep{pinte08} and IRAS\,04158+2805 \citep{glauser08}. In the former
two cases, these studies have unambiguously shown that the dust
population is stratified, possibly indicating that dust settling is
already occurring. Increasing the number of disks studied in such
detail is necessary to disentangle individual peculiarities from
genuine trends associated with disk evolution.

{\hv} is a triple system located in the Taurus star-forming region. It
consists of a 550\,AU wide pair whose optically brightest component is
itself a tight (10\,AU) visual binary \citep{simon96}. Spectroscopic
and photometric measurements revealed that this subsystem does not
currently experience accretion nor does it show infrared excess. They
further establish an age of about 2\,Myr for the system
\citep{white01}. The third component of the system, {\hvc}, is much
fainter yet bluer than {\hvab}. While \citet{magazzu94} first thought
that this source was an Herbig-Haro object, \citet{woitas98} later
proposed that {\hvc} is a normal M0 T\,Tauri star surrounded by an
opaque edge-on disk similar to that found in HH\,30 by
\citet{burrows96}. Subsequent high resolution imaging confirmed this
hypothesis \citep{monin00}. \citet{stapelfeldt03} produced the first
model of high resolution 0.8 and 2.2\mic\ scattered light images of
{\hvc}, finding that dust properties similar to those of interstellar
dust grains can account for these images. Their 0.8\mic\ image also
revealed the presence of a roughly spherical envelope, producing a
symmetric halo that is more extended than the disk itself, that is
likely the remnant of the core from which the system formed.

As a consequence of their particular viewing geometry, edge-on
protoplanetary disks offer a unique opportunity to determine their
geometry and dust content. As such, they may be the best candidates to
study vertical stratification in protoplanetary disks. They also are
comparatively easy targets for high-angular resolution instruments
since the contrast requirement is strongly relaxed. On the other hand,
they are challenging from the modeling point of view because of the
difficulty of accurately solving for radiative transfer including
anisotropic scattering in high optical depth regions. Previous
modeling efforts of edge-on protoplanetary disks have therefore
focused on interpreting one type of observation at a time
\citep{burrows96, stapelfeldt98, stapelfeldt03, cotera01, wood02,
  watson04}. While these studies proved highly valuable to constrain
some of the disk properties, no self-consistent model was used to
model all data at once, leaving unexplained contradictions.

Our objective in this work is to perform a global analysis of the
{\hvc} disk, combining scattered light images, millimeter
interferometric data and the overall SED into a single fit. We present
a series of new observations of the system in Section~\ref{sec:obs}
and discuss our empirical results in Section~\ref{sec:results}. In
Section~\ref{sec:models}, we present radiative transfer models of the
{\hvc} disk and discuss their implications in
Section~\ref{sec:discus}. Section~\ref{sec:concl} summarizes our
results.

\section{Observations}
\label{sec:obs}

\subsection{Adaptive optics near-infrared imaging}

\subsubsection{1--2\mic\ imaging}

On 2002 November 24, we observed {\hv} using the NAOS adaptive optics
system and the CONICA instrument \citep{lenzen03,rousset03} installed
on the Yepun 8.2\,m-Unit Telescope at ESO's Very Large Telescope, as
part of the NAOS Guaranteed Time Observing program. We used the
0\farcs0133 pixel scale for the $J$ ($\lambda_0=1.27$\mic,
$\Delta\lambda=0.25$\mic) and $H$ ($\lambda_0=1.66$\mic,
$\Delta\lambda=0.33$\mic) images and the 0\farcs0270 plate scale for
the $K_{\rm s}$ ($\lambda_0=2.18$\mic, $\Delta\lambda=0.35$\mic)
images. {\hvab}, a $m_{\rm V}=14.5$ source, was used as adaptive
optics guide star with the visible wavefront sensor. From narrow-band
observations of single stars throughout the night, the measured FWHM
of point-like sources is about 0\farcs10, 0\farcs07 and 0\farcs08 at
$J$, $H$ and $K_{\rm s}$ band, respectively. These images, similarly,
have Strehl ratios of approximately 5\%, 25\% and 40\%.

We obtained deep images in which {\hvab} is saturated, as well as
shallow images with an additional neutral density filter and shorter
exposures to record unsaturated images of the primary. Total
integration times for the long exposures of 750s, 480s and 165.5s
were recorded at $J$, $H$ and $K_{\rm s}$, respectively, split in 8 to
20 dithered independent images. Total integrations times for the
shallow images were 100s, 50s and 45s, respectively at $J$, $H$ and
$K_{\rm s}$. For each sequence of images, a sky was estimated by
medianing all images, which was then subtracted from each image prior
to cosmetic cleaning, which included bad pixels and cosmic ray
correction, and flat-fielding. All resulting images were then
shift-and-added to produce final images.

On 2002 November 26, we re-observed {\hv} with NAOS and CONICA using
the 0\farcs0270 pixel scale, this time using the 1\farcs4-diameter
coronagraph mask to block out the starlight from {\hvab} and record
deeper exposures on {\hvc}. Three 120s $K_{\rm s}$ band images were
recorded. A sky was subtracted off each image before they were
cosmetically cleaned and flat-fielded. Finally, they were averaged to
yield the final coronagraphic image. No point source is detected in
the 30\arcsec\ field-of-view, preventing us to estimate the achieved
FWHM and Strehl ratios.

\subsubsection{3--5\mic\ imaging}

On 2002 December 13, we observed {\hv} using the NIRC2 camera
(P.I. K. Matthews) installed behind the Keck\,II adaptive optics
system \citep{wizi00} to record images using the $L'$
($\lambda_0=3.78$\mic, $\Delta\lambda=0.70$\mic) and $M_{\rm s}$
($\lambda_0=4.67$\mic, $\Delta\lambda=0.24$\mic) filters and the
0\farcs00996 plate scale \citep{ghez08}. {\hvab} was used as a Natural
Guide Star (NGS) for the adaptive optics system. To reduce thermal
background and provide a more symmetric and smoother PSF, we used the
``inscribed circle'' pupil stop, resulting in an effective primary
mirror diameter of 9m. At $L'$, four images, consisting of 300 coadded
0.181s individual integrations each (for a total of 217.2s on source
integration), were acquired with the sources dithered on-chip between
images. A fifth image was acquired with all stars moved out of the
detector field-of-view (10\arcsec) to estimate the sky level. At
$M_{\rm s}$, 500 individual 0.150s integrations were coadded using a
reduced 6\arcsec-wide field-of-view, necessary to avoid background
saturation on the intense thermal background. A total of 21 such
images were obtained with the stars moved about in the available
field-of-view, representing a total integration time of 1575s. The
corresponding sky was estimated by median-combining all frames.

On 2004 November 03, we re-observed {\hv} with NIRC2, this time using
the newly available Laser Guide Star (LGS) module on the Keck\,II
adaptive optics system \citep{vandam06,wizi06}. The laser system was
run at 6 Watts output power, which produced a guide star with an
equivalent magnitude of $m_{\rm V}=9.9$, and {\hvab} was used as the
tip/tilt correction point source. To take full advantage of the better
image quality, the ``largehex'' pupil mask, which does not block any
section of the primary mirror, was used throughout the
observations. {\hv} was imaged in the $L'$ filter with an 0.2s
integration coadded 100 times. This observing cycle was repeated 27
times, with the system alternatively located in opposite detector
quadrants, providing a total integration time of 540 secs. For each
image, the subsequent one, with the star located in the opposite
quadrant, was used as a sky frame.

All datasets were reduced using a similar strategy to that used in
processing the NACO images \citep[see also][]{duchene04}. First of
all, the sky thermal emission was subtracted. These subtracted images
were then flat fielded and had any bad pixels interpolated over. In
the case of the LGS dataset, a residual sky level in the frames was
measured by taking the median value of the quadrants which do not have
a star in the field. This value (typically corresponding to 0.05\% of
the initial sky value) was then subtracted off the cleaned images. All
cleaned images in a given dataset were then median-averaged. We
measured FWHMs of 0\farcs09 and 0\farcs11 for point-like sources
observed before/after {\hv} at $L'$ and $M_{\rm s}$ for the NGS
dataset and 0\farcs08 at $L'$ for the LGS dataset. We estimated the
corresponding Strehl ratios to be about 70\% at $L'$ (both in NGS and
LGS mode) and 80\% at $M_{\rm s}$, using the ``Strehl meter'' tool
developed by the Keck
observatory\footnote{http://www2.keck.hawaii.edu/optics/aochar/Strehl\_meter2.htm}.

Given the high thermal background at $L'$ and $M_{\rm s}$, the
resulting signal-to-noise on {\hvc} is limited, about 15, 20 and 5 in
the peak pixel at $L'$ (NGS, LGS) and $M_{\rm s}$, respectively. To
enhance detection without altering the intrinsic spatial resolution of
the images, we smoothed them by using a running 2-pixel-radius
($\approx$0\farcs02) median-filtering circular mask, and rebinned them
by a factor of 3 in both directions, resulting in an approximate final
sampling of 0\farcs03/pixel that still oversamples the resolution of
the datasets.

\subsection{Sub-millimeter and millimeter imaging}

\subsubsection{1.3 and 2.8\,mm interferometric imaging}

On 2005 February 26, we observed {\hv} with IRAM's 6-antenna Plateau
de Bure Interferometer \citep[PdBI,][]{guilloteau92} in the 6Bp
configuration, with baselines ranging from 71\,m to
331\,m. Simultaneous 110\,GHz (2.76\,mm) and 230.5\,GHz (1.31\,mm)
observations of {\hv} were recorded in double-sideband mode with a
bandpass of 640\,MHz at each frequency. {\hv} was observed
alternatively with phase calibrators 0415+379 and 0528+134 through an
entire 11h-transit, resulting in beam sizes of
1\farcs1$\times$0\farcs9 (along position angle 38\degr) and
2\farcs1$\times$1\farcs6 (along position angle 62\degr) at 1.3 and
2.8\,mm, respectively. The average weather conditions resulted in
r.m.s. phase noises on the order of 15--40\degr\ at 2.8\,mm and
8--15\degr\ at 1.3\,mm (using self-calibration from the 2.8\,mm data),
equivalent to an atmospheric ``seeing'' on the order of 0\farcs3 and
0\farcs6 at 1.3 and 2.8\,mm, respectively. The absolute pointing
uncertainty is on the order of 0\farcs1--0\farcs2. The quasars
NRAO\,150 and 3C\,273 were used as absolute flux calibrators,
resulting in a 10\% uncertainty on all quoted fluxes. The data were
reduced using the GILDAS package and selecting individual baseline
visibilities for which the phase noise was less than 40\degr\ and flux
variations were less than 20\% based on calibrators
measurements. Simultaneous observations of the $^{12}$CO 2--1
transition (230.538\,GHz rest frequency) with a 20\,MHz bandpass were
obtained to probe the gaseous component of the disk. Data reduction
followed the same method as the continuum data, providing a
3-dimension reconstructed datacube with a 0.1\,km.s$^{-1}$ spectral
resolution.

We obtained follow-up observations of {\hv} with the 15-antenna CARMA
array in its C configuration, with baselines ranging from 24 to
300m. On 2008 May 6 and 9, we tuned the receivers to a central
frequency of 110\,GHz (2.76\,mm), while on 2008 May 29 we tuned them
to 90\,GHz (3.33\,mm). The total continuum bandwidth is 2.8\,GHz,
split in 6 separate bands. Observing conditions were average and parts
of the observations had to be flagged out because of poor phase
coherence. Overall, the useful integration times on {\hv} were 3h45 at
110\,GHz and 2h30 at 90\,GHz. Observations of {\hv} were interleaved
with pointings at 3C\,111 and J\,0530+135 which served as phase
calibrators; flux calibration was performed by observing a planet
(Uranus, Neptune) at the beginning of each track. The systematic
uncertainty in CARMA's absolute flux scale is $\sim$20\% (W. Kwon,
priv. comm.). The data were reduced using the MIRIAD software
package. The final 110\,GHz map corresponds to the combination of both
observing periods, and is characterized by a beam size of
1\farcs8$\times$1\farcs4 (along position angle 116\degr); the 90\,GHz
map has a beam size of 3\farcs1$\times$1\farcs7 (along position angle
124\degr).

\subsubsection{350\mic\ single dish mapping}

On 2008 January 28, we observed {\hv} at 350\mic\ with the
32$\times$12 SHARC-II bolometer array \citep{dowell03} installed at
the Caltech Submillimeter Observatory. The array was scanned at an
8\farcs2/s rate over a 60\arcsec$\times$ field centered on {\hv}. This
scanning was repeated until a total integration of 600\,s was
achieved. Ceres and HL\,Tau, two secondary flux calibrators, were
observed immediately after with total integration times of 180\,s and
300\,s, respectively. The fluxes of both these sources are known to
within 10\%; their measured fluxes agreed to within this
uncertainty. Conditions were excellent for submillimeter observations
($0.013\leq\tau_{\rm 225\,GHz}\leq0.036$). All data were reduced
using the CSO-developed CRUSH software, using a 6\arcsec\ smoothing
for {\hv} (a 4\arcsec\ smoothing was used for the flux calibrators),
to produce a final image that has an effective resolution of
approximately 10\arcsec.

\section{Observational results}
\label{sec:results}

\subsection{{\hv} as a triple system}

At all wavelengths, our adaptive optics images clearly reveal the
typical morphology of {\hvc} as an opaque edge-on disk, namely two
parallel, horizontal nebulae separated by a dark lane
(Figure~\ref{fig:newimgs}). They further show {\hvab} to be
systematically extended along the north-western direction, although
the pair is barely resolved due to its tight separation (see
Figure~\ref{fig:orbit}). Here we discuss the relative astrometry of
all 3 components.

Focusing first on the wide pair {\hvc}--{\hvab}, we estimate its
separation and position angle based on the location of the centroid of
both components. Despite saturation of {\hvab} in some images, all
images yield consistent estimates (see Table~\ref{tab:binary}), with
an average separation of 4\farcs04$\pm$0\farcs02 and a position angle
of 44\fdg6$\pm$0\fdg6. There is marginal ($\approx3\sigma$) evidence
for a variation of the binary position angle as a function of
wavelength, though this could be due to underestimated uncertainties
on the absolute orientation of the detectors. We do not find any
significant trend as a function time at our level of precision. Based
on the proper motion of {\hvab} measured by \citet{ducourant05}, the
projected separation and position angle of the pair would have changed
in the almost 5 years between the observations of
\citet{stapelfeldt03} and ours by a total of 0\farcs12 and 2\fdg2,
respectively, if {\hvc} had not been comoving with {\hvab}. This can
be rejected at the $\approx5\sigma$ level in our datasets, and is
further confirmed by the relative astrometry obtained at earlier
epochs by \citet{simon92} and \citet{woitas98}. We therefore conclude
that {\hv} is a common proper motion pair with a projected separation
of 565\,AU, making it a bona fide triple system.

To obtain reliable relative photometry and astrometry for the tight
{\hvab} pair, we used PSF-fitting. We choose to analyze the short $H$-
and $K_{\rm s}$-band VLT exposures and both $L'$-band Keck images, as
they offer the best image quality and the most favorable ratio between
binary separation and achieved resolution. For the VLT $H$-band and
Keck $L'$-band images, we searched for adequate PSFs among images of
single stars obtained on the same night as our {\hv} images. For the
VLT $K_{\rm s}$-band, we used a set of single stars from observations
taken with the same set-up presented in \citet{duchene07}. PSF-fitting
was performed using IRAF's {\it daophot} package. The resulting
relative astrometry and photometry is listed in
Table~\ref{tab:binary}.  The VLT and Keck NGS data have been taken
within a few weeks of each other, and no measurable orbital motion is
expected over such a short timescale. Averaging these three datasets,
we find a separation of 0\farcs0595$\pm$0\farcs0025 and a position
angle of 312\fdg5$\pm$1\fdg8. The Keck LGS data, obtained two years
later, yields a relative astrometry that is marginally consistent
($2.5\sigma$) with this estimate. The projected separation in our
images appears to be 4--5$\sigma$ smaller than earlier measurements
\citep{simon96,monin00}, but there is no significant change in
position angle as a function of time (see
Figure~\ref{fig:orbit}). This relative displacement is most likely due
to orbital motion. The observed plane-of-the-sky velocity of {\hvab},
$\approx$1.5\,km.s$^{\rm -1}$, is roughly one order of magnitude too
low considering the 10\,AU projected separation of the binary compared
to other T\,Tauri binaries \citep[e.g.,][]{ghez95}. The number of
resolved measurements and the total amplitude in orbital motion are
insufficient to attempt an orbital fit for {\hvab} at this
point. Nonetheless, the unexpectedly low measured orbital velocity
suggests a highly elliptical orbit observed around apoastron passage,
or a large out-of-the-plane separation which, combined with the nearly
radial observed motion, would in turn imply that the orbital plane is
almost perpendicular to the plane of the sky. Monitoring of the system
in the next few years will help disentangle these two possibilities.

Our PSF fitting also yields relative photometry for the {\hvab}
binary. Considering the near-simultaneous $H$, $K_{\rm s}$ and NGS
$L'$ images, we find evidence that {\hvb} is somewhat redder than
{\hva} in the near-infrared (see Table~\ref{tab:binary}). In the
framework in which none of the components possesses circumstellar
material, this is an indication that {\hvb} is cooler than {\hva},
consistent with it being fainter. Surprisingly, \citet{simon96} found
a flux ratio in the visible that is closer to unity than our
near-infrared measurements. Temporal variability of either component
could be an explanation, although we note that both our $L'$ flux
ratios are consistent with one another despite being taken two years
apart. Further monitoring is required to better understand the
intrinsic colors of both components.

\subsection{The {\hvc} circumstellar disk}

\subsubsection{Scattered light images}

The near-infrared images presented here are of higher spatial
resolution than those presented in \citet{monin00} and
\citet{stapelfeldt03}, which both had a 0\farcs13 resolution, and are
comparable to the $HKL'$ Subaru images of \citet{terada07}. Our
observations extend the wavelength coverage of {\hvc} to $M_{\rm s}$
for the first time and, combined with the HST F814W image of
\citet{stapelfeldt03}, offer an almost uniform spatial resolution
(0\farcs07--0\farcs11, or 10--15\,AU) view of the disk over the entire
0.8--4.7\mic\ range. This is critical to conduct unbiased studies of
the wavelength dependence of the disk images.

To quantify the basic morphological properties of {\hvc}, we adopt the
following method. At each wavelength, we first estimate the position
angle of the dark lane, $PA_{\rm disk}$, as the average of the
position angles of each nebulae, as determined from fits of elliptical
Gaussian intensity profiles. Averaging all estimates, we find a
position angle of 108\fdg3$\pm$0\fdg4, which we take as the
orientation of disk midplane. The dark lane width, $d_{\rm neb}$, is
measured as the projection of the vector joining the light centroid of
the two nebulae on the disk minor axis. Peak-to-peak flux ratios
($FR_{\rm peak}$) are readily estimated, whereas integrated flux
ratios ($FR_{\rm int}$) are obtained by summing the flux within areas
that encompass all pixels whose surface brightness is at least 5\%
(20\% for the $M_{\rm s}$ image) of the peak surface brightness in the
image. Finally, we measure the total extent of the disk along its
major-axis, $w_{\rm 5\sigma}$, defined by the horizontal extent of the
contour at the 5$\sigma$ noise level. We note that the spherical halo
identified by \citet{stapelfeldt03} dominates this measurement in the
F814W image. All of these quantities are given in
Table~\ref{tab:observables}.

Several wavelength-dependent features can be noted in our images of
{\hvc}. As Figure~\ref{fig:contours} illustrates, $d_{\rm neb}$
decreases by about 30\% from 0.8 to 4.7\mic. We further find that both
$FR_{\rm peak}$ and $FR_{\rm int}$ smoothly decline towards longer
wavelengths without ever reaching unity; rather, both flux ratios
actually plateau longwards of 2.2\mic. Finally, the intensity profile
of the bright nebula along the disk major axis is remarkably invariant
from 0.8 to 4.7\mic\ (see Figure~\ref{fig:profiles}). The measured
FWHM is about 0\farcs35 at all wavelengths, i.e., a factor of 3--4
larger than the resolution of our images. This achromatic behavior is
therefore not influenced by observational limitations but is rather
intrinsic to the disk.

One main feature of the {\hvc} disk that was pointed out by
\citet{stapelfeldt03} is the lateral asymmetry: the northeastern
(bright) nebulae extends further on one side while the southwestern
(faint) nebula extends further on the other side. This can be readily
seen in Figure~\ref{fig:contours}. In addition to this low-intensity
asymmetry, the location of the nebulae centroids is also non
symmetric, suggesting a global asymmetry: the line joining the two
centroids is misaligned by about 10\degr\ with respect to the disk
symmetry axis as determined by the orientation of the dark lane. We do
not find any significant variation of the asymmetry with wavelength
nor time within our uncertainties.

\citet{stapelfeldt03} also discussed narrow structures, which they
dubbed ``rays'', extending along the disk major axis beyond the
main/bright nebula which they associate to the disk, with total extent
up to 1\farcs4 in their optical images. We can track scattered light
up similar distances at both $K_{\rm s}$ and $L'$ in our images, a
regime in which the halo is undetected. Although the exact nature of
these features remains uncertain, this suggests that these ``rays''
trace the surface of the disk instead of being a mere shadow of the
disk on the spherical halo. In that case, the actual disk size may be
larger than the 50\,AU assumed in the past.

\subsubsection{Thermal emission regime}

Both our PdBI and CARMA observations have a sufficient spatial
resolution to disentangle the wide {\hv} pair. In all maps, a single
source is detected, at the location of {\hvc}. {\hvab} appears to have
no significant emission at millimeter wavelengths with 3$\sigma$ upper
limits in our PdBI data of 6\,mJy and 1.5\,mJy at 1.3 and 2.8mm,
respectively.

At 1.3\,mm, {\hvc} is spatially resolved in our PdBI data, with a
sharp decrease in correlated flux at the longest baselines along the
position angle of the disk as defined from the scattered light images
(see Figure~\ref{fig:fcor}). We therefore fit an elongated Gaussian
profile to the visibilities, assuming that its minor axis is
unresolved which is consistent with the data. This yields a total flux
density of 49.5$\pm$1.8mJy, a FWHM of 1\farcs17$\pm$0\farcs08 along
the major axis, at PA 111$\pm$3\degr\ East of North, in excellent
agreement with the position angle derived from the scattered light
images. This analytical fit is shown in Figure~\ref{fig:fcor};
deviations between observations and our simple fit result from the
fact that the distribution of surface brightness in the disk is not a
perfect Gaussian. The measured 1.3\,mm flux agrees well with the
previous single-dish measurement by \citet{beckwith90}. At 2.8\,mm,
the source is unresolved with the PdBI, with a flux density of
7.1$\pm$0.5mJy. The CARMA observations do not resolve the disk either,
and we extracted point-source fluxes of 8.0$\pm$2.1\,mJy at 2.8\,mm
and 3.8$\pm$0.9\,mJy at 3.3\,mm. Flux calibration uncertainties of
10\% and 20\% must be added for the PdBI and CARMA observations,
respectively.

A single point-like source is detected in our 350\mic\ CSO map of
{\hv} with a flux of 0.370$\pm$0.030\,Jy; a 10\% uncertainty for flux
calibration must be quadratically added. The resolution of this
dataset, roughly 10\arcsec, does not allow us to spatially resolve the
wide pair. However, taking advantage of the strong detection of the
system ($\approx20\sigma$), we use the nearly simultaneous observation
of the bright source HL\,Tau with the same set-up to determine an
accurate astrometric positioning \citep{mundy96}. The 350\mic\ source
is found to be located at (04h38m35.50s, +26\degr10\arcmin40\farcs6,
J2000) with an uncertainty of about 1\arcsec\ in both directions. This
is in excellent agreement both with the position of the lone
millimeter source in our PdBI and CARMA maps and with the location of
{\hvc} in optical and near-infrared images. We therefore conclude that
{\hvc} is the dominant source of emission at 350\mic\ and assign all
of the observed flux to that component.

\subsubsection{Gas emission}

As shown in Figure~\ref{fig:co}, $^{12}$CO 2--1 line emission is
detected at the position of {\hvc} in our PdBI observations. The
integrated spectrum shows two distinct peaks, with a near-zero trough
separating them. The trough in the middle of the line is likely due to
contamination by large-scale emission from the surrounding molecular
cloud that is filtered out by the interferometer. \citet{mizuno95}
detected $^{13}$CO emission in the 5.5--7\,km~s$^{-1}$ from the
molecular cloud at the location of {\hv}, consistent with this
interpretation. We further find that the blue- and red-shifted parts
of the line emission are spatially distinct and symmetric about the
continuum peak. The two line emission peaks are separated by about
1\farcs5, or 200\,AU, spatially and 4\,km.s$^{-1}$ in
velocity. Finally, we note that the CO emission appears to extend
beyond both the millimeter continuum and the scattered light images,
suggesting that there is more to the system than meets the eye in
continuum datasets. 

\subsubsection{The SED of \hvc}

To draw a complete picture of the {\hvc} disk, we compiled its SED by
combining published fluxes with our new measurements. From the same
references (see below), we also built the SED of {\hvab}
(Figure~\ref{fig:sed}), which is well reproduced from the optical to
the mid-infrared by a 3600\,K, $\log g = 4.0$, NextGen model from
\citet{baraffe98} assuming $A_{\rm V}=1.75$\,mag, in agreement with
previous spectrophotometric estimates of the stellar properties of
that component \citep{kenyon95,white01}.

Constructing the SED of {\hvc} is a delicate matter because of the
large scatter in its published optical and near-infrared photometry
\citep[see also][]{monin00,stapelfeldt03}. Figure~\ref{fig:sed}
includes all published data and illustrates this variability. To
construct a ``representative'' SED for {\hvc}, we elected to aim for
smoothness. For instance, we adopt the most recent $KLN$ photometry
from \citet{mccabe06}, which matches well with the {\it Spitzer}/IRAC
data from \citet{hartmann05}. We discard the $N$ band measurements
from \citet{woitas98} as they appear to overestimate the flux of both
components by about 1\,mag. Based on their agreement at $K$ band with
the McCabe et al. photometry, we then adopted the $JH$ fluxes from
\citet{woitas98}. Finally, we adopted the visible photometry from
\citet{magazzu94} which offers a smoother extension of the SED to
short wavelengths than that of \citet{stapelfeldt03}, although it was
obtained almost a decade earlier. Apart from the $K$ band measurement
by \citet{simon92}, our selection is equivalent to consistently
selecting the brightest measurement available at every wavelength. At
longer wavelengths, we used the MIPS 24\mic\ and 70\mic\ detections
and a 170\mic\ upper limit from the {\it Spitzer} Taurus Legacy Survey
(Rebull et al., submitted). At 24\mic\, the astrometric accuracy is
good enough to identify the detected source with {\hvc} with little or
no contribution from {\hvab}. In the longest wavelength regime, we
adopt fluxes measured from our data and from \citet{andrews05}. The
resulting SED, the most complete to date for an edge-on disk system,
is shown as filled diamonds in Figure~\ref{fig:sed} and presented in
Table~\ref{tab:photom}.

The SED of {\hvc} can be characterized as a ``double hump'' (see
Figure~\ref{fig:sed}), which is typical of other edge-on disks
\citep[e.g.,][]{strom94,stapelfeldt99,wood02}. The short wavelength
hump is dominated by photons emitted from the central star and the
innermost region of the disk and scattered to the observer at the
outer radius of the disk, whereas the long wavelength part represents
the disk thermal emission propagated through the mostly optically thin
disk to the observer. The wavelength of the turnover between the two
humps is driven by the total column density of dust along the line of
sight to the central star. We find this turnover to be between 8 and
11.8\mic\, although a more precise estimate would require analysis of
a mid-infrared spectrum of the source. At longer wavelengths,
\citet{andrews05} noted that {\hvc} has an abnormally flat
850\mic--1.3\,mm slope. With our new observations, it appears that the
850\mic\ flux is indeed too low compared to extrapolation from longer
wavelength fluxes. Limiting our analysis to the 1.3--3.3\,mm regime,
we find a spectral index of $\alpha_{\rm mm}=2.5\pm0.2$, where $F_\nu
\propto \nu^{\alpha_{\rm mm}}$. This value is on the low end of the
distribution observed for other protoplanetary disks
\citep{beckwith91,mannings94,natta07}.

\subsubsection{Qualitative interpretation}

Here we interpret some of the observational results listed above to
make qualitative inferences about the {\hvc} system. We will then
revisit in a quantitative way these conclusions on the basis of the
radiative transfer modeling presented in Section~\ref{sec:models}.

The $\gtrsim$1\,mag variability of {\hvc} in the near-infrared is
larger than has been recorded for other T\,Tauri stars in the past
\citep{eiroa02,alves08}. It is therefore unlikely that this
variability results from intrinsic fluctuations in the emission of the
star and inner disk. A plausible alternative scenario could invoke a
self-shadowing pattern in the disk that moves as a result of
differential Keplerian rotation. In this framework, the structure
responsible for the lateral asymmetry in the images would be located a
few AU away from the central star to account for the absence of
variability over the course of 2 years. This could be linked to the
observed asymmetry in the scattered light images. Such a phenomenon
has been documented in the HH\,30 edge-on disk \citep{watson07}. One
possibility is that the disk is warped as a consequence of the
gravitational forces exerted by {\hvab}.

The wavelength dependent features of the scattered light images inform
us about the absorption and scattering properties of the dust grains
in the disk. For instance, the decreasing $d_{neb}$ towards longer
wavelengths is a consequence of the declining dust opacity
\citep{stapelfeldt03,watson04}. The fact that the intensity profile
along the bright nebula does not change with wavelength is suggestive
of a dust population whose phase function is equally anisotropic at
all wavelengths. A dust population similar to that of the interstellar
medium, whose scattering asymmetry parameters drops rapidly beyond
1\mic\ \citep[e.g.,][]{weingartner01}, seems inconsistent with our
observations of {\hvc}. We note, however, that the flux ratio between
the nebulae is not constant across the 0.8--4.7\mic\ range, even
though it is also believed to be driven by the scattering asymmetry.

The slope of the SED of an edge-on disk between the near- and
mid-infrared is a function of the disk geometry and of the grain size
distribution and composition in the disk scattering surface. In their
analysis of the ``Flying Saucer'' edge-on disk, \citet{pontoppidan07}
concluded that the shallow 2--10\mic\ slope in that system implies the
presence of a ``significant amount of 5--10\mic\ grains'' in that
disk. Such grains are necessary to produce a high albedo in the
mid-infrared. While it is tempting to apply this argument to {\hvc},
uncertainties about the disk geometry prevents from reaching a firm
conclusion just yet.

If the disk were optically thin to its own emission in the millimeter
regime, the observed spectral index could be readily converted into an
opacity power law index of $0.5\pm0.2$, characteristic of evolved dust
grain populations that extend up to a few millimeter in size
\citep{mannings94,natta00,andrews05}. However, because the disk is
very compact and viewed almost exactly edge-on, the optically thin
assumption may be incorrect, so that grain growth cannot be claimed on
this sole basis in this particular system. If the disk were indeed
optically thick, one would expect the millimeter spectral index to
steepen at longer wavelengths. We find
$\alpha_{\mathrm{1.3-2.7mm}}=2.5\pm0.2$ and
$\alpha_{\mathrm{2.7-3.3mm}}=3.4\pm0.6$, a difference that is not
statistically significant. Flux measurements at even longer
wavelengths are ultimately needed to conclude on this possibility.

In any case, there are several independent hints of the presence of
grains at least a few microns in size, and maybe up to millimeter
sizes, in the {\hvc} disk. However, none of these inferences are
robust enough as they depend on some assumptions about the disk
geometry, among other factors. The goal of the modeling presented in
the next section is to simultaneously fit for the disk structure and
the dust properties so as to solve for these ambiguities.

In the absence of a gas tracer whose emission from the cloud is
negligible, we do not perform a complete Keplerian rotation analysis
of our $^{12}$CO observations. Qualitatively, {\it the measured
  amplitude of the velocity and positional offsets are consistent with
  an 0.5--1$M_\odot$ central star} (see
Figure~\ref{fig:co}). Observations in other molecular tracers are
necessary to go beyond this simple analysis, however. We also note
that the larger outer disk radius suggested by the gas emission is not
a unique property of the {\hvc} disk. Indeed, this phenomenon is
rather common among protoplanetary disks
\citep[e.g.,][]{pietu05,isella07,panic09}, although few objects have
been mapped at very high resolution in both the continuum and line
emission. \citet{hughes08} have recently demonstrated that this can be
reproduced if the surface density profile of disks is tapered, rather
than sharply truncated, outside of a certain radius, for
instance. Based on the gas emission alone, the disk outer radius might
be as large as 100\,AU, twice the estimate from the continuum and
scattered light images. Since our entire analysis focuses on the dust
component of the disk, we do not consider such large disk radii in our
modeling.

\section{Panchromatic Modeling}
\label{sec:models}

The objective of this section is to compare our broad set of
observations of {\hvc} to predictions of a radiative transfer code in
order to constrain the main properties of the disk. Since this is the
first attempt at simultaneously reproducing scattered light images,
thermal emission maps and the SED of {\hvc}, our approach consists in
considering a simplified parametrized disk structure in an effort to
search for a model that would represent a reasonable global fit to,
rather than an exact representation of, all observations of this disk.

\subsection{Radiative transfer models: overall framework}

We use the MCFOST radiative transfer code \citep{pinte06} which
computes synthetic observables, such as SEDs and images, by
propagating photon packets through the disk. Scattering, absorption
and reemission by dust grains are taken into account following the Mie
theory valid for homogeneous spherical grains. We assume that the dust
grains are at the local thermal equilibrium with the surrounding
radiation field throughout the disk. Synthetic temperature maps, SEDs
and images are computed simultaneously at all inclinations, allowing
us to fit for that parameter as well. For each model, the temperature,
SEDs and images computation are performed using typically 1, 4 and 32
million packets, respectively. The disk is assumed to be passive,
i.e., its only source of heating is radiation from the central star.

The disk geometry is described by its inner and outer radii, $R_{\rm
  in}$ (which we fix at 0.15\,AU) and $R_{\rm out}$, as well as by two
independent power laws. The surface density profile of the disk
follows $\Sigma(r) = \Sigma_0 (r/r_0)^\alpha$. Vertically, we assume
that $\rho(r,z) = \rho_0(r) \exp[-z^2/(2H(r)^2)]$, appropriate for a
vertically isothermal, non-self-gravitating disk in hydrostatic
equilibrium. Finally, we adopt a flaring law for the scale height,
namely $H(r) = H_0 (r/r_0)^\beta$. The dust content of the disk is
characterized by a power law size distribution, $\mathrm{d}N(a)
\propto a^{-p}\mathrm{d}a$ from $a_{\rm min}$ to $a_{\rm max}$. We
adopt the optical properties of the commonly-used ``astronomical
silicate'' mixture from \citet{draine03} with $p=3.7$ and $a_{\rm
  min}=0.03$\mic. In all our models, we adopt a default distance of
140\,pc, consistent with the system's parallax estimated by
\citet{bertout06}. The central star is described by an effective
temperature of 3800\,K based on it spectral type
\citep{woitas98,white01,appenzeller05}, and we use the corresponding
$\log g = 4.0$ NextGen synthetic spectrum \citep{baraffe98}. Although
$R_\star$ is difficult to constrain in {\hvc} since we cannot measure
the total bolometric luminosity of the object, we consider it as a
free parameter since there is no simple method to fix it a priori.

\citet{stapelfeldt03} showed that adding an envelope to the system
yields a much improved fit to the visible scattered light images of
{\hvc}, especially to account for the roughly spherical halo that is
detected well above the disk midplane. To implement this, we add a
spherically symmetric envelope to our model, which is characterized by
$R_{\rm in}=1$\,AU and $R_{\rm out}=85$\,AU \citep[maximum extent of
the ``rays'' identified by][]{stapelfeldt03} to mimic the halo seen in
the HST images. For this envelope, we use a total dust mass of $5
. 10^{-8}\,M_\odot$, a radial density profile $\rho(r)\propto
r^{-0.5}$ and interstellar-like dust grains ($a_{\rm max}=1$\mic\ and
$p=3.7$). The optical depth through the envelope at $\lambda=0.5$\mic\
is $\tau\approx0.5$, i.e. the maximum acceptable considering the
morphology of the visible images of the disk. While a spherical
envelope has little physical relevance, as its free-fall time would be
much shorter than the system age, we do not explore more sophisticated
envelope models (e.g., rotation-supported). Indeed, we consider our
approach as a reasonable ``placeholder'' to avoid systematically poor
$\chi^2$ values for our model scattered light images. Images of {\hvc}
show that the envelope is only significantly detected shortward of
1\mic, anyway, and even then the region of highest signal-to-noise
ratio, which contributes most to the total $\chi^2$ is dominated by
scattering off the disk. We therefore do not expect our simplistic
parametrization to induce any strong bias on our results. Finally, we
note that the asymmetry of the disk seen in scattered light image
produces such strong deviations from our simple model that attempting
to adjust the envelope profile is of little interest at this point.

\subsection{Parameter space exploration}

\subsubsection{Strategy and modeling grid}

While we have set a number of parameters of the model, we still have
to deal with an 8-dimensional parameter space: the total dust mass in
the disk ($M_{\rm dust}$), $R_{\rm out}$, $H_0$ (defined at
$r_0=50$\,AU), $\beta$, $\alpha$, $a_{\rm max}$, $R_\star$, and the
inclination to our line of sight ($i$, where $i=0$\degr\ corresponds
to a face-on disk) are all free parameters. When computing synthetic
SEDs, we also added a foreground extinction $A_{\rm V}$ following the
interstellar extinction law which we considered as an additional free
parameter, with values ranging from 0 to 5\,mag. This extinction
represents attenuation by material located around or between the {\hv}
system and us and does not include attenuation by the edge-on disk and
spherical envelope themselves.

For all model parameters except the inclination, we adopt a coarse
sampling that encompasses reasonable estimates either from the
previous modeling by \citet{stapelfeldt03} and \citet{andrews05} or
our own estimates above (see Table~\ref{tab:models}). For instance, we
chose a range of values for $a_{\rm max}$ that extends from 1\mic\ to
1\,cm. As shown in Figure\,\ref{fig:opacity}, $a_{max}=1$\mic\ results
in an opacity law that is in reasonable agreement with the measured
interstellar extinction law through the mid-infrared. By extension, we
will refer to this model as ``interstellar dust''. The only parameter
for which we use a large number of possible values is the
inclination. MCFOST produces synthetic disk observables at all
inclinations at once. The final images and SED are stored in several
``inclination bins'' which are equally spaced in cos($i$), appropriate
for a random orientation prior. By using 91 independent inclination
bins in all our simulations, we obtain a 0\fdg6 sampling in the
vicinity of the perfectly edge-on geometry.

Exploring the parameter space with such a coarse sampling is
computationally manageable, although it comes at the cost of failing
to find a model that fits perfectly all data. Our original goal was to
use this grid to identify a small region of the parameter space that
produced a good fit to all observations and to run a second, finer
grid in that smaller regions. However, as we discuss below, our
analysis revealed that no single region of the parameter space could
be selected that way, as different observations point towards
different parts of the parameter space. We discuss possible avenues
for improvement for future modeling efforts in
Section~\ref{sec:discus}. With 8 free parameters in our model grid,
we have computed over a million independent models, which required
about 126,000\,h of CPU time on a 400-processor cluster, 92\% of which
was devoted to the synthetic SED calculation.

Because we chose a fixed value for $R_{in}$, the temperature at the
inner edge of the disk varies from 800 to 1300\,K in the
grid. Similarly, the temperature at the outer edge ranges from 15 to
30\,K. The radial dependence of the dust temperature in the midplane
outside of $\sim 0.25$\,AU closely follows a power law, $T\propto
r^{-q}$, whose index is in the range $q=0.4$--0.6 and mostly depends
on the amount of flaring in the disk. The dust temperature at a given
location depends primarily on $L_\star$, hence $R_\star$ and, to a
lesser extent, on $a_{\rm max}$ (via the total dust opacity) and other
geometric parameters. Our modeling strongly favors
$R_\star=3\,R_\odot$ (see below), for which the range in maximum
temperature in the disk is 1100--1350\,K, close to the sublimation
temperature of dust grains. While setting $R_{in}$ on the basis of
this sublimation temperature for all models may be more physically
relevant \citep{muzerolle03}, it is a computationaly expensive
iterative process, and we believe that it would not dramatically
change our conclusions.

We decided to model 5 independent observables of {\hvc}, which we
consider as representative of the entire dataset available to us: its
broadband SED (Table~\ref{tab:photom}), the F814W, $H$ and $L'$ (LGS)
scattered light images, and our IRAM 1.3mm visibilities as a function
of projected baselines. Our choice of wavelengths for the scattered
light images to fit for is a trade-off between considering as wide a
wavelength range as possible while securing data that has high enough
spatial resolution and signal-to-noise per pixel. The latter criterion
led us to discard our new $M_{\rm s}$ image in this modeling. In any
case, a model that reproduces the F814W, $H$ and $L'$ images
reasonably well is likely to also reproduce scattered light images at
other wavelengths throughout the entire 0.8--5\mic\ range. MCFOST
synthetic scattered light images were computed as monochromatic images
at the effective central wavelength of each filter and with a
2\arcsec\ total field-of-view. The thermal emission from both the disk
and the envelope is neglected in computing these images, as most of
the emission arises from the inner most regions, which is virtually
unresolved as seen from the disk outer edge, where scattering towards
the observer occurs. As discussed below, we only aim at reproducing
the morphology and we have checked that this morphology is virtually
unchanged if we take into account the disk emission, which otherwise
induces a heavy computational cost. In the millimeter regime, we
computed 1.3\,mm thermal emission maps with a very fine spatial
sampling to avoid aliasing in the Fourier transform (0\farcs05/pixel).

\subsubsection{Goodness-of-fit estimates}

While we are interested in finding the best possible model to account
for the observations of {\hvc}, we also aim at determining the range
of acceptable models around it. For this purpose, we adopt a Bayesian
inference method \citep[e.g.,][]{lay97,akeson02,pinte08}, in which
each model is assigned a probability that the data are drawn from the
model parameters. In cases where the prior has a uniform probability
distribution, as is the case here either on a linear, cosinus-like or
logarithmic scale, this probability is $P = P_0\exp(-\chi^2/2)$, where
$\chi^2$ is the reduced chi square. The normalization constant, $P_0$,
is chosen so that the sum of the probabilities over all models in the
grid is unity. Once this is done for all models in our grid, we can
derive the probability distribution for a given parameter by
marginalizing the 8-dimensional probability hypercube against the
other 7 dimensions.

The computation of $\chi^2_{\rm SED}$ has 12 degrees of freedom. To
take into account the noise intrinsic to Monte Carlo simulations, we
ran 10 independent realizations of one of the best-fitting model and
quadratically added the resulting standard deviation to the
observational uncertainties. For the number of packets we used, this
Monte Carlo noise is about 10\% in the optical/near-infrared and
(sub)millimeter regimes but reaches about 40\% in the mid-infrared
where photons have the hardest time escaping their deeply embedded
emission region. Even though it is likely that this overestimates the
uncertainties for non-edge-on models (for which photons escape more
easily to the observer), it is of little importance since these models
have very poor $\chi^2$ values to start with, given that they do not
produce the characteristic double-hump SED. In computing $\chi^2_{\rm
  SED}$, we chose to fit for $\ln (\nu F_\nu)$ instead of $\nu F_\nu$,
as this better handles the case of datasets that are dominated by
calibration uncertainties (i.e., all measurements beyond 100\mic),
which are multiplicative rather than additive in nature.

For scattered light images, we perform a pixel-by-pixel computation,
though with a few adjustments. First of all, we resampled the $H$ band
image to a pixel scale of 0\farcs027 for a higher signal-to-noise
without compromising the spatial resolution. All synthetic images were
then convolved with the same PSFs as used for the deconvolution of
{\hvab} (and a TINYTIM-generated F814W
PSF\footnote{http://www.stsci.edu/software/tinytim/tinytim.html}). We
then aligned the model and empirical images using cross-correlation in
the Fourier domain to determine the offsets with which the model
images are best aligned with the observed ones. We further normalized
all images to a peak value of unity since our goal is to reproduce the
morphology of the images, the flux being fitted for in the
SED. Finally, to avoid including many pixels where both the data and
model are indistinguishable from zero and that would artificially
improve our $\chi^2$, we selected to use square binary masks
encompassing the area where there is significant flux from
{\hvc}. These masks are 2\arcsec, 1\farcs6 and 1\farcs4 on a side at
F814W, $H$ and $L'$, respectively, resulting in 2017, 3713 and 2201
degrees of freedom. Tests conducted without these masks showed that
this does not affect the ranking of the models from best to worse
while ensuring that the best $\chi^2$ are not artificially low, which
would result in too relaxed constraints. Similar to the SED $\chi^2$
computation, we estimated Monte Carlo noise maps by computing 10
realization of the best-fitting model at all three wavelengths. For
the number of packets we used here, relative uncertainties range from
less than 2\% at the image peak to 8--12\% at the disk's outer edge.

At 1.3\,mm, the available data are a series of correlated fluxes as a
function of projected baselines. To reduce the number of visibilities
to compute for each model and to improve their signal-to-noise ratio,
we first flagged out all correlated fluxes whose uncertainty was
larger than 0.1\,Jy, i.e., twive the total flux from the source
itself. We then averaged the correlated fluxes in the (u,v) plane
using a 50\,m bin size. This yields 60 independent measurements, or 52
degrees of freedom, at 1.3\,mm. The model images were first rotated so
that the disk midplane lies at a position angle of 108\fdg3, and
convolved with a 0\farcs3 Gaussian ``seeing''. After padding the
images with zeros to obtain the adequate total field of view, we
computed their Fast Fourier Transform to obtain a set of synthetic
correlated fluxes. As for the treatment of the scattered light
described above, here we do not wish to fit for the total flux in the
1.3\,mm map but for its morphology. Therefore, we convert the
correlated fluxes into visibilities using the observed 49.5\,mJy flux
for the input data and the measured total flux in each map for the
models. The 1.3\,mm synthetic maps are essentially noiseless and we
neglect the Monte Carlo noise in this regime.

\subsection{Modeling results}

\subsubsection{Best-fitting models}

The $\chi^2$ values for the models that best fit each observation of
{\hvc} are listed in Table~\ref{tab:chi2}. The combination of
observational and numerical uncertainties used in computing the
goodness-of-fit of a given model is unlikely to follow exactly a
normal distribution. Caution should therefore be used when
interpreting individual $\chi^2$ values as they could be
systematically biased. Nonetheless, the ranking of the models is
likely correct and the model with the absolute lowest $\chi^2$ value
should be close to what would have been the best possible fit within
the explored parameter space in the absence of non-Gaussian noise.

First of all, we note that the best reduced $\chi^2$ for the SED, $H$-
and $L'$-band images fit are in the 3.7--4.9 range. Considering the
intrinsic asymmetry of the disk and the coarse sampling of the
parameter space, we consider that these observations are well
reproduced. While the intrinsic structure of protoplanetary disks is
likely to be much more complex than the power law used here, the
simplified parametrization adopted here appears reasonable and can
therefore provide valuable insight about the disk properties. The fact
that the best $\chi^2_{\rm F814W}$ is only 11.5, i.e., much worse than
the other two scattered light images, results in part from its higher
signal-to-noise ratio which amplifies any departure from the
models. In addition, the relative undersampling of the WFPC2 image
results in a poorer ability to register images, thereby globally
increasing all values of $\chi^2_{\rm F814W}$. Lastly, we note that
the range of $\chi^2_{\rm 1.3mm}$ in our entire grid is very narrow,
about a factor of 2 from best to worst, implying that this observation
of {\hvc} is only mildly constraining for our model. This was to be
expected considering that the disk is only marginally resolved in our
PdBI data.

Figure~\ref{fig:fit_sed} compares the observed SED of {\hvc} to
several ``best'' models. The best overall model has a foreground
extinction of $A_{\rm V}=1$\,mag, marginally lower than the extinction
we estimated for {\hvab}. Both the model that best fits the F814W
image and the best overall model with $a_{\rm max}=1$\mic\ (i.e.,
interstellar-like dust) fail badly at reproducing the photometric
data, with $\chi^2_{\rm SED}$ of 139.9 and 20.7,
respectively. Although these models produce a millimeter flux that is
a factor of at least a few lower than the observed flux, it is
interesting to note that they yield a shallow millimeter spectral
index as a result of their high optical depth. Indeed, all models that
adequately reproduce at least one of the {\hvc} observations have
$\tau_{\rm 1.3mm}\gtrsim15$ in the disk midplane. This is a clear
reminder that millimeter spectral indices can be strongly affected by
optical depth effects in the case of edge-on disks. Another
shortcoming of an interstellar-like dust model, as discussed by
\citet{pontoppidan07}, is that it yields a near- to mid-infrared slope
that is far too steep compared to observations if they are required to
produce the right flux ratio between the two humps of the SED. This is
a result of the vanishingly small mid-infrared albedo of any dust
model that does not include grains of several microns in
size. Therefore, the SED of {\hvc} points to the presence of at least
intermediate-size grains in the disk.

Let us now consider the model scattered light
images. Figure~\ref{fig:fit_imgs} compares the observed images to
several key models. In addition to the factors listed above, the
poorer fit to the F814W image likely arises from our simplistic
treatment of the spherical halo, which is too bright at large
distances above the disk midplane in the models. The models that best
reproduce individual images each provide a good fit to the main
features of the {\hvc} disk, such as the distance and flux ratio
between the two nebulae. The model that offers the best trade-off
between the three scattered light images nicely reproduces the
wavelength dependency of $d_{\rm neb}$ but has roughly achromatic flux
ratios between the nebulae. This is likely a consequence of the fact
that this model, which has $a_{\rm max}=1$\,cm, has an almost
wavelength-independent scattering phase function, which is required to
reproduce the intensity profiles shown in
Figure~\ref{fig:profiles}. Interestingly, while the best model for
each individual image corresponds to $R_{\rm out}=75$\,AU, the best
scattered light and best overall model both have an outer radius of
50\,AU. Again, the disk asymmetry prevents us from unambiguously
estimating the disk outer radius based on the images only. Finally, we
note that the model that best fits the SED is a very poor fit to the
scattered light images ($\chi^2$ ranging from 34.3 to 68.1). The
inclination of that model is 76\fdg3, giving the observer an almost
grazing angle line-of-sight, resulting in a counternebula that is
almost undetectable and an almost point-like appearance at $L'$ which
does not match the observed image at all.

Although the 1.3\,mm emission map is approximately reproduced by
virtually all models in our grid, it is noteworthy that the
best-fitting models have a face-on inclination. This stems from the
fact that the disk, if viewed edge-on, is unresolved along its minor
axis given the beam of our PdBI observations. However, our data show a
slight drop of the correlated fluxes at the longest baselines along
that position angle. Therefore, our modeling prefers a lower
inclination in order to explain this feature. Since we know that the
{\hvc} disk is indeed edge-on, this may indicate that the atmospheric
phase noise increases with baseline in a way that differs from our
Gaussian ``seeing'' approximation. Alternatively, it could be that the
intrinsic thermal emission from the spherical envelope is stronger
than in our model, which only contains small dust grains that are
inefficient emitters at this regime. While this could in principle
induce a bias in our modeling, we consider this effect as negligible
considering that this dataset is the least important in constraining
the disk model.

\subsubsection{Bayesian analysis}

Figure~\ref{fig:bayes_all} shows the inferred probability
distributions for each of our 8 free parameters, after marginalization
against all 7 other parameters. Fitting for each scattered light image
separately yields probability distributions that are similar to each
other and we only show the probability distribution for fitting all
three images at once, for clarity purposes. The only differences
between fits to individual images, associated with inclination and
maximum grain size, are discussed below. The results from the fit to
the F814W image tends to drive the results when grouping all three
scattered light images because the other images are comparatively
easier to reproduce, as shown by the lower best $\chi^2_{\rm H}$ and
$\chi^2_{\rm L'}$ compared to the best $\chi^2_{\rm F814W}$.

We remind the reader that the derived probabilities that a given
parameter takes a certain value are only valid within the framework of
our modeling and are based on our approximate treatment of Monte Carlo
noise in the simulations. Nonetheless, they represent a more reliable
metric in our analysis than $\chi^2$ values for individual models and
they therefore better highlight how each observation informs our model
of the disk. We note that the best individual model (shown as filled
diamonds in Figure\,\ref{fig:bayes_all}) has parameter values that
correspond in most cases to the most likely value as determined from
the Bayesian method, or to a value whose global probability is at
least 20\%. This demonstrates that noise in both data and model is
well-behaved, providing support to the results of the Bayesian
analysis.

The scattered light images place some constraints on $R_{\rm out}$,
$a_{\rm max}$, $i$ and, to a smaller extent, $M_{\rm dust}$. Since we
have used flux-normalized images, $R_\star$ remains
unconstrained. Similarly, $\alpha$, $\beta$ and $H_0$ are not well
constrained as a result of trade-off between these parameters and
$M_{\rm dust}$. A scale height of 4--5\,AU is preferred when fitting
only the F814W but it is not conclusive and the other images have much
flatter probability distributions for $H_0$. Our modeling of the
scattered light images finds $R_{\rm out}=50$\,AU as the most likely
value, although an outer radius of 75\,AU cannot be formally excluded
considering the disk asymmetry and total extent. The probability
distribution for $M_{\rm dust}$ peaks at the lowest end of the range
sampled here, and $P(M_{\rm dust}\leq10^{-4}M_\odot)\approx75$\%. It
is worth noting that the probability distributions based on fitting
each image individually are almost flat, but combining all three in a
single fit yields a significant constraint as it helps solving for
some of the ambiguities between parameters. Similarly, we find that
$P(a_{\rm max}\leq100\mu m)\approx90$\% when fitting all three images
at once, but there is an underlying trend as a function of
wavelength. Indeed, $P(a_{\rm max}\leq1\mu m)$ drops from 45\% (most
likely) at 0.8\mic, to 20\% at 1.6\mic\ and to 7\% (least likely) at
3.8\mic, respectively. This gradual and smooth evolution with
wavelength is probably indicative of a real physical
phenomenon. Finally, the combined fit to the scattered light images
suggests an inclination\footnote{Uncertainty ranges are defined by the
  34-percentile on each side of the most likely value.} of
82\fdg7$^{+0\fdg6}_{-1\fdg9}$. However, there is a marginal
(2$\sigma$) trend as a function of wavelength: the most likely
inclinations are 80\fdg8$^{+1\fdg3}_{-1\fdg9}$, 84\fdg6$+/-$2\fdg5 and
85\fdg9$^{+1\fdg3}_{-2\fdg5}$ for the F814W, $H$ and $L'$ images,
respectively. This trend stems from the inability of our model to
produce a wavelength-dependent flux ratio between the nebulae without
generating chromatic variations of the intensity profile along the
major axis. This may indicate that our dust model does not possess the
adequate chromatic behavior.

Considering now the SED, our modeling mostly provides constraints on
the same parameters as the scattered light images. The SED fits favor
the presence of large grains and a high total dust mass, with
$P(a_{\rm max} > 100\mu m) \approx 85$\% and $P(M_{\rm dust} > 10^{-4}
M_\odot) \approx 85$\%. Both conditions are necessary to produce
sufficient fluxes at the long wavelength end of the SED. While
interstellar-like grains can produce the observed spectral index, all
models with such a dust population fall short in the 1--3\,mm range by
at least an order of magnitude. While SED fitting is generally not
sensitive to the outer radius of a disk, here we find that a small
outer radius is strongly preferred: $P(R_{\rm out}\leq50\,AU) /
P(R_{\rm out}\geq75\,AU) \approx 4$. This is because the relative
height of the two humps in the SED and the depth of the trough that
separates them is directly influenced by the disk geometry, via the
total optical depths along our line of sight to the star on one hand
and from the star to the upper layers of the outer disk on the other
hand. Finally, the modeling of the SED yields a best fit inclination
of 75\fdg0$^{+4\fdg5}_{-3\fdg9}$. The fact that the SED provides a
weaker constraint on the disk inclination than the scattered light
images is typical of all disk analyses. Nonetheless, the particular
viewing angle of edge-on disk allows for a reasonably narrow range of
inclination to be defined, and it is reassuring to note that the
difference between the inclinations derived from the scattered light
images and the SED is at the insignificant 1.6$\sigma$ level.

The fit to the 1.3\,mm yields much weaker constraints on the model
parameters: all associated probability distributions are consistent
with being flat except for $R_{\rm out}$, $\alpha$ and, to a smaller
extent, $i$. As discussed above, this is a result of the fact that the
disk is only marginally resolved in our PdBI data. To best reproduce
the observed visibilities, the model favors a face-on geometry, a
large outer radius and a flat surface density that places a lot of
mass, hence of millimeter emission, at large radii. As far as
inclination is concerned, the inferred most likely value is
0$^{+50}$\degr, illustrating the weak preference for a face-on
geometry. The constraints on both $R_{\rm out}$ and $\alpha$ are much
stronger with a ratio of the most- to least-likely value of at least
4:1.

The strength of our modeling approach is to be able to fold all
independent datasets in a single coherent fit, which both yields much
sharper constraints and a view of the intrinsic contradictions between
observations. For some parameters, such as $\beta$, $H_0$ and
$R_\star$, the combined fit provides better constraints than any
individual observation as a consequence of the fact that each
observation is associated to a different set of ambiguities. We thus
find that the disk is significantly flared ($\beta\gtrsim1.15$), that
its scale height at 50\,AU is 5\,AU at most, and that
$R_\star\sim3\,R_\odot$, which implies
$L_\star\sim1.7\,L_\odot$. Assuming that the accretion luminosity is
negligible, this is consistent with a 1\,Myr 0.7--1\,$M_\odot$ central
star \citep{baraffe98}, sightly larger than the mass estimated for
{\hva} \citep{white01}. Although uncertainties are large, this is in
rough agreement with the dynamical mass that we have estimated from
the rotating CO disk.

In the case of $R_{\rm out}$ and $\alpha$, the constraints set by the
millimeter mapping are superseded by those from the scattered light
images and the SED because the former is comparatively too easy to
reproduce in our grid. While a $1/r$ surface density profile is
preferred in our model, all other values tested in our grid have
non-negligible probabilities. Similarly, our overall fit favors
$R_{\rm out}=50$\,AU only by a factor of 2 over the larger
radius. Excluding the 1.3\,mm map from our analysis increases the
preference for the smaller outer radius to a factor of 4:1,
however. Future higher resolution millimeter observations are needed
to better constrain both of these parameters. Also, the final
probability distribution for the disk inclination is largely driven by
the scattered light images. Our global fit suggests an inclination of
82\fdg1$^{+2\fdg5}_{-1\fdg9}$.

The most interesting results of our analysis concern $a_{\rm max}$ and
$M_{\rm dust}$, for which scattered light images and SED lead to
contradictory predictions (see Figure~\ref{fig:bayes_all}. The latter
favors large grains and a high dust mass whereas the former suggests
small grains and a low dust mass. Our global fit, which aims at
finding the best possible trade-off between all constraints points to
a very high dust mass ($M_{\rm dust}\geq10^{-3}\,M_\odot$) and an
intermediate maximum grain size ($a_{\rm max}\approx10$\mic). Since
both of these values are essentially rejected ($P\lesssim10$\%) by at
least one of the observations, this ``best'' model should not be
considered as a good fit. Rather, this apparent contradiction between
the various observations, which probably mirrors the trend in
preferred $a_{\rm max}$ as a function of wavelength for the scattered
light images, indicates that our model is too simplistic in at least
some aspects. Possible explanations are explored in
Section~\ref{sec:discus}.

\subsection{Comparison to previous modeling}

There are three published models of the {\hvc} disk. In their
discovery study, \citet{monin00} derived a disk inclination of
$\approx84$\degr\ based on a single scattering model. That inclination
has later been confirmed by subsequent models of the disk, including
the present work. The second, most sophisticated to date, model of the
disk was put forth by \citet{stapelfeldt03} who analyzed scattered
light images at 0.8 and 2.2\mic. More recently, \citet{andrews05}
derived a total disk mass for {\hvc} based on its sub-millimeter flux
although they did not conduct a complete SED fit for this source. Here
we compare our model results to these previous efforts.

The model constructed by \citet{stapelfeldt03} was based on the same
F814W image we have used here, and a lower quality 2.2\mic\ adaptive
optics image that was characterized by a poorer resolution (0\farcs13
compared to 0\farcs08 for our VLT/NAOS image) and a smaller achieved
Strehl ratio. Their conclusions are similar to ours in several
ways. First of all, they found that it is easier to fit for the $K$
band image, as demonstrated by the lower achieved $\chi^2$
value. Second, fitting for the envelope-free $K$ band image alone
points towards an inclination that is closer to edge-on than if the
F814W image is included in the fit. Conversely, when fitting both
images simultaneously, they also note that they do not succeed in
reproducing the wavelength dependence of the flux ratio between the
nebulae. 

If we temporarily focus on the F814W image alone, which is the most
constraining dataset in their fit, our Bayesian analysis yields
$a_{\rm max}=1$\mic, $M_{\rm dust}=10^{-5}M_\odot$ and $H_0=4$\,AU and
$i=80$\fdg8. These values are consistent with those of
\citet{stapelfeldt03} within our uncertainties. The derived
inclination, somewhat further away from edge-on, may be explained by
our different treatment of the spherical halo, which plays a
non-negligible role in shaping up the morphology of {\hvc} at
0.8\mic. Finally, we note that our simultaneous fit to the F814W, $H$
and $L'$ images favors $a_{\rm max}=1$\mic. This model has a ratio of
opacity of $\kappa_{0.8\mu\mathrm{m}} / \kappa_{2.2\mu\mathrm{m}}
\approx 3.3$ and scattering asymmetry parameters of
$g_{0.8\mu\mathrm{m}}=0.6$ and $g_{2.2\mu\mathrm{m}}=0.55$. These
properties are very similar to those derived by \citet{stapelfeldt03},
namely $\kappa_{0.8\mu\mathrm{m}} / \kappa_{2.2\mu\mathrm{m}} \approx
3.5$ and $g_{0.8\mu\mathrm{m}}=0.65$.

Overall, we find that our fit to scattered light images is in good
agreement with the previous modeling effort for this source, despite
differences in strategy and parameter space covered. However, the
models that best reproduce the F814W image produce millimeter fluxes
that are one to two orders of magnitude too low in the millimeter
regime, in apparent contradiction with the conclusion of
\citet{stapelfeldt03}, who found agreement within a factor 2. The
explanation for this discrepancy is that these authors assumed a ratio
of dust opacity of $\kappa_{0.8\mu\mathrm{m}} /
\kappa_{1.3\mathrm{mm}} = 6.10^3$. However, with our assumed dust
composition, this ratio is about 7.10$^4$ for $a_{\rm max}=1$\mic,
accounting for most of the difference in predicted millimeter flux
between the two models.

\citet{andrews05} presented submillimeter observations of {\hvc},
which we have included in our fit. Because the source showed an
unusual spectral index, they did not attempt to fit a full-fledged
model but rather used an empirical recipe to convert the measured
850\mic\ flux into a disk mass. They derived $M_{\rm dust}=2
. 10^{-5}\,M_\odot$, almost two orders of magnitude lower than the
value we have derived here based on either the SED or global fit. The
overall shape of the SED for {\hvc} suggests that the 850\mic\ flux is
underestimated, possibly a factor of 3 or so, accounting for part of
this discrepancy. In addition, the empirical law that
\citet{andrews05} used was mostly derived from sources which are not
edge-on and for which optical depth effects are negligible. However,
the model defined by the combination of each of the most likely
parameter value has $\tau_{1.3\mathrm{mm}}\gtrsim100$ in the disk
midplane. We believe that this factor is the most important in
explaining the difference in the disk masses inferred here and by
\citet{andrews05}.

\subsection{Shortcomings of the model}

While our model has been successful at reproducing the SED of {\hvc},
as well the morphology and several key chromatic behavior of its
scattered light images, it falls short in several aspects which are
worth exploring. As expected, a significant shortcoming of our model
is its built-in axisymmetric assumption, which prevents us from
finding perfect matches to any of the scattered light images. Since
the nature of the asymmetry can only be speculated upon, no clear path
to resolution can be provided until further monitoring clarifies its
origin. This peculiarity is however not sufficient to account for the
poor $\chi^2_{\rm tot}$ of our overall best fitting model. The most
glaring limitation of our model is the fact that it cannot account
simultaneously for the SED on one hand and the scattered light images
on the other. This fact, which most likely is independent of the
axisymmetry of the model, implies that more complexity has to be
included in the model.

The physical properties of dust grains are subject to
improvement. Based on our modeling experience, the necessary
requirements for an improved dust model in {\hvc} would be an
interstellar-like opacity law and achromatic scattering phase function
from 0.8 up 5\mic\ and an optical-to-millimeter opacity ratio
$\lesssim10^4$. For one, one could devise a dust model that contains a
fraction of C-rich grains to better conform to interstellar
abundances. Departures from a simple power law size distribution are
also likely. Such a phenomenon has been observed in the interstellar
medium \citep{kim94,weingartner01} and multi-modal distributions are
predicted by dynamical models of grain growth, migration and
fragmentation in protoplanetary disks \citep[e.g.,][]{dullemond05,
  laibe08, zsom08}. Another direction to explore is the possibility of
non-spherical grains. Recently, \citet{kimura03} successfully
reproduced the wavelength-independent anisotropic scattering observed
in cometary dust by considering aggregates of small particles. Since
the dust in {\hvc} also has achromatic scattering properties,
aggregate grains are a candidate that deserves consideration in the
future. In this context, it is interesting to note that fractal grains
are expected to have a much flatter optical-to-millimeter opacity law
\citep{wright87}, akin to our models with $a_{max}\geq100$\mic, which
are preferred in our analysis.

Since it appears difficult to simultaneously match all of these
constraints with a single dust model, it will likely be necessary to
drop the assumption that the dust properties are uniform throughout
the disk. Since each type of observation probes a different region of
the disk, {\it decoupling the dust properties in various regions of
  the disk would likely provide sufficient leeway.} In this picture,
the disk would be quite massive and would contain large grains, up to
millimeter-sized particles, in the midplane in order to explain the
observed SED. However, grains in the surface layers would not exceed a
few microns in size to explain the scattered light images. To limit
the number of free parameters, we have only considered fully mixed
models here but we suggest that this hypothesis be further studied in
future modeling of {\hvc}.

Future modeling efforts on {\hvc} should consider these various
possibilities in improving on the model presented here. A robust
strategy to move forward could consist in conducting detailed modeling
of a particular type of observations, such as that conducted by
\citet{watson04} on the scattered light images of HH\,30 for
instance. These specific analyses would allow one to extract the
maximum amount of information from each observation and to feed it
into a refined global model.

\section{Discussion}
\label{sec:discus}

\subsection{Global properties of the {\hvc} disk}

Our objective in this work was to conduct the first ``global'' fit to
the structure and dust content of the {\hvc} disk, one of the few
objects in which such an effort is conducted in a self-coherent
way. Although some uncertainty remains, our analysis has led to some
robust conclusions about the {\hvc} disk. For instance, the total mass
of the disk must be at the high end of the range probed here. Even in
the presence of large grains, which are efficient emitters in the
millimeter regime, the total dust mass has to be on the order of
$M_{\rm disk}\sim10^{-3}\,M_\odot$. Indeed, such a high dust mass is
necessary to account for the observed fluxes and relatively high
optical depths up to 3.3\,mm. Assuming a canonical 100:1 gas-to-dust
mass ratio, this implies that the disk is quite massive compared to
the central star ($M_{\rm disk}/M_\star\gtrsim0.1$). This is at the
high end of the distribution observed for disks surrounding T\,Tauri
stars \citep{andrews07}. It is plausible that the disk is only
marginally stable, in which case the presence of spiral density waves
could be responsible for the observed asymmetry and/or variability of
the system.

One respect in which the {\hvc} disk is special is its compact radius,
a probable consequence of tidal forces induced by {\hvab}. With
$R_{\rm out}=50$\,AU and $\alpha=-1$, we derive a total surface
density of 2800 g~cm$^{-2}$ 1\,AU away from the star, comparable to
the surface density inferred for the early Solar Nebula as well as for
extra-solar planetary systems, albeit with a possibly flatter surface
density profile \citep{hayashi81,kuchner04}. Indeed, the surface
density at large radii in the disk, almost 60\,g.cm$^{-2}$ at 50\,AU,
appears substantially higher than those derived for other
protoplanetary disks \citep{dutrey96,kitamura02}. {\it The disk around
  {\hvc} is a clear example of the fact that binary systems can host
  circumstellar disks massive enough to form planets for timescales of
  several million years}. Indeed, HV Tau may well be a prototype for
the earliest evolutionary stages of field stars that are found to host
both an extrasolar planet and at least one stellar companion
\citep[e.g.,][]{eggenberger07,mugrauer09}.

As a test of the self-consistency of our model, it is interesting to
compare the disk scale height we have derived, $H_0\leq5$\,AU, to the
simple assumption of vertical hydrostatic equilibrium that is embedded
in our parametric disk structure. Assuming molecular hydrogen, the
scale height can be written as $H_0^{\rm hydro} = 3.4 (T/20K)^{1/2}
(R/50AU)^{3/2} (M_\star/M_\odot)^{-1/2}$\,AU. Our best global model
has a midplane temperature of 19\,K at 50\,AU, so that our upper limit
on $H_0$ implies $M_\star\geq0.44M_\odot$. This is not a stringent
constraint on the mass of the central star, but we note that this
lower limit does not violate the mass inferred from the spectral type
and luminosity of {\hvc} nor the rough kinematic estimate we derived
from the rotating CO disk. Both estimates are in the 0.7--1\,$M_\odot$
range, for which the derived scale height would be 3.3--4\,AU at
50\,AU, which is consistent with our modeling. We therefore conclude
that the outer disk can reasonably be described by a disk in
hydrostatic equilibrium. In addition, {\it we suggest that {\hvc} may
  be the most massive component in the triple system.}

Finally, while large uncertainties remain regarding the dust
properties, {\it it is clear that dust grains much larger than 1\mic\
  are found in at least some parts of the disk}. This could be
evidence for grain growth in the disk, although it is plausible that
growth to a few microns may have predated the formation of the disk
itself \citep{mccabe03}. Indeed, analyses of extinction laws have
already pointed to the presence of small quantities of grains a few
microns in size in the interstellar medium and even more so in
molecular clouds \citep{kim94, weingartner01, indebetouw05,
  flaherty07}. On the other hand, growth to mm-sized grains, which
seems supported by the analysis of the SED of {\hvc}, has most likely
occurred in the disk itself. If the disk is indeed vertically
stratified and no such grains are found in the disk surface, it
remains unclear whether this is because of a very efficient settling
process or a mere consequence of the difficulty to grow such large
particles in the lower density regions of the disk.

\subsection{Comparison to other circumstellar disks}

As outlined in the introduction, evidence for grain growth and dust
stratification in protoplanetary disks has been mounting over the last
two decades. It is therefore not surprising that our modeling of the
{\hvc} disk calls for both processes. However, one must keep in mind
that a given observation only probes a limited region of the disk. For
instance, mid-infrared emission is dominated by emission from the
inner few AUs at most, providing no information on dust stratification
in the outer disk. In addition, there is an intrinsic observational
bias in all studies whereby only a small range of grain sizes, those
with the largest effective cross-section, actually contribute to the
observed fluxes. Therefore longer wavelength observations tend to
systematically call for larger grain sizes, even if one limits the
analysis to scattered light images
\citep[e.g.,][]{mccabe03}. Similarly, the mid-infrared silicate
features arise from micronic and submicronic grains, providing no
constraint on much larger particles. To draw a {\it complete} picture
of dust properties throughout a given disk, it is therefore necessary
to gather wide and homogeneous datasets, as we have done here.

As we have already mentioned, few objects have been studied in as much
detail as our present analysis of {\hvc}. A clear dichotomy in terms
of grain size was also found in the case of IRAS\,04302+2247
\citep{wolf03} but the interstellar-like dust grains favored by the
scattered light images were located in the relatively massive envelope
surrounding this embedded object and not in the disk itself. Vertical
stratification was demonstrated in the massive GG\,Tau ring
\citep{duchene04}, although the available evidence only applies to
grain smaller than 10\mic. The HK\,Tau\,B edge-on disk also appears to
have a vertically stratified structure \citep{duchene03}, although a
self-consistent modeling of its SED, thermal emission maps and
scattered light images is still required to make direct comparison
with {\hvc}. On the other hand, \citet{glauser08} were able to
reproduce a wide range of observations of IRAS\,04158+2805 without the
need to include vertical stratification. A similar conclusion was
recently reached by \citet{sauter09} for CB\,26. Overall, these
detailed analyses confirm the general trends outlined above that both
grain growth and stratification are common, but not ubiquitous, in
protoplanetary disks. They further offer the unique opportunity to
test the detailed physics of dust evolution, such as the amount of
dust stratification or its radial dependency \citep[e.g.,][]{pinte07}.

An interesting head-to-head comparison can be drawn between {\hvc} and
IM\,Lup, for which \citet{pinte08} recently conducted a similar
panchromatic modeling analysis, including scattered light images,
millimeter emission maps and the system's SED. The central stars have
the same spectral type, and similarly high disk masses were inferred,
making these two systems very similar. Contrary to the conclusion
reached here however, \citet{pinte08} were able to place stringent
constraints on all disk parameters and one may wonder what the reasons
for this difference are. For one, the IM\,Lup disk is not seen
edge-on, which allows the authors to fix some of the model parameters
(e.g., $R_\star$, dust composition, $M_d$) in an unambiguous way. The
smaller dimensionality of the parameter space to explore allowed
\citet{pinte08} to sample it more finely. Another key feature of the
IM\,Lup disk is that it is much less compact than {\hvc} and is well
resolved at millimeter wavelengths, providing more stringent
constraints on the model. Last but not least, a qualitative analysis
of the SED of IM\,Lup readily demonstrated the need for a stratified
structure in the disk, which introduces an extra degree of freedom
that we could not afford in our analysis of {\hvc}. Nonetheless, our
study has shown that {\it significant constraints on the disk
  properties can be obtained in an edge-on geometry}. No global
modeling of HH\,30 or HK\,Tau\,B, other prototypical edge-on disks,
are yet available, but conducting such efforts should yield some
interesting results and are needed to move forward in our
understanding of protoplanetary disks.

\section{Conclusion}
\label{sec:concl}

We have obtained new 1--5\mic\ adaptive optics images of the {\hv}
triple system at the VLT and Keck observatories, including the first
4.8\mic\ scattered light image of the edge-on {\hvc} disk.  All of our
images reveal a steady lateral asymmetry in the disk that indicates a
departure from pure axisymmetry in the disk. This could be related to
the known variability of {\hvc}. We extact precise relative astrometry
from our new images and find that {\hvab}--C constitutes a common
proper motion pair. We also resolved the tight binary system {\hvab}
and found a surprisingly slow orbital motion compared to its projected
separation, probably due to orbit eccentricity and/or a large
deprojected physical separation. We have also obtained the first
spatially resolved 1.3\,mm continuum and $^{12}$CO 2--1 millimeter
maps of the {\hvc} disk with IRAM's PdBI interferometer. The continuum
emission is resolved along the same position angle as the scattered
light images, as expected from dust thermal emission. The CO map shows
evidence of Keplerian rotation about an 0.5--1$M_\odot$ central
star. We also completed the SED of that component with new {\it
  Spitzer}/MIPS 24, 70 and 160\mic\ observations as well as
observations at 350\mic, 2.8 and 3.3\,mm at the CSO, PdBI and CARMA
facilities. The 1--3\,mm spectral index of {\hvc} is relatively flat
($\alpha_{\rm mm}=2.5\pm0.2$), indicative of the presence of
millimeter-sized grains in the disk and/or high optical depth even at
millimeter wavelengths.

To interpret these data along with previously published HST images of
{\hvc}, we have computed a grid radiative transfer models that
produced synthetic disk observations. While we can reproduce most
observational properties of {\hvc}, a Bayesian analysis of our model
grid reveals that the SED and scattered light images of {\hvc} provide
mutually exclusive constraints. As found previously, the scattered
light images are best fit with a relatively small total mass and
interstellar-like dust properties. Fitting the SED, however, requires
almost two orders of magnitude more mass as well as a grain size
distribution that extends to millimeter sizes. The mismatch between
the two sets of constraints reveals an intrinsic shortcoming of our
parametrized model, which assumes perfectly mixed dust. Indeed, the
small maximum dust grain size inferred from fitting the scattered
light images is impossible to reconcile with the long-wavelength end
of the SED. In turn, this suggests that both grain growth and vertical
stratification are present in the {\hvc} disk. While these phenomena
have been suggested for other protoplanetary disks in the past, {\hvc}
is one of a handful of objects that have been studied in sufficient
details to eventually provide a complete picture of these processes.

Future observations of {\hvc} can help refine the model proposed
here. For instance, high-resolution (0\farcs2 or better) millimeter
mapping of the disk would better resolve it and provide more stringent
constraints. In particular, obtaining observations at even longer
wavelengths, 7\,mm or 1.3\,cm, would probe a regime in which the
optical depth through the disk is much smaller and enable a more
direct interpretation in terms of disk properties. On the longer run,
mapping of the disk with ALMA in the submillimeter regime and in
scattered light at 8--10\mic\ with the next generation of large
ground-based telescopes will help bridge the gap between the
scattering and thermal emission regimes at a roughly constant spatial
resolution. In particular, very high resolution maps with ALMA may
resolve this and other disks along the vertical axis, in order to
probe their vertical stratification and, ultimately, to outline an
evolutionary sequence.

\acknowledgments

We are indebted to all members of the GEODE group for many fruitful
discussions about modeling of edge-on protoplanetary disks, and in
particular to Marshall Perrin for developing some of the model
analysis tools used in this project. We are grateful to Darren Dowell
for his help with the astrometric calibration of our CSO data and to
the CARMA staff for conducting the observations presented in this
paper. The work presented here has been funded in part by National
Science Foundation Science and Technology Center for Adaptive Optics,
managed by the University of California at Santa Cruz under
cooperative agreement No. AST - 9876783, by the Programme National de
Physique Stellaire of CNRS/INSU (France) and by the Agence Nationale
de la Recherche through contract ANR-07-BLAN-0221. C. P. acknowledges
the funding from the European Commission's Seventh Framework Program
as a Marie Curie Intra-European Fellow (PIEF-GA-2008-220891). The
authors wish to acknowledge the contribution from Intel Corporation,
Hewlett-Packard Corporation, IBM Corporation, and the National Science
Foundation grant EIA-0303575 in making hardware and software available
for the CITRIS Cluster which was used in producing some of the model
computations presented in this paper. Support for CARMA construction
was derived from the Gordon and Betty Moore Foundation, the Kenneth
T. and Eileen L. Norris Foundation, the Associates of the California
Institute of Technology, the states of California, Illinois, and
Maryland, and the National Science Foundation. Ongoing CARMA
development and operations are supported by the National Science
Foundation under a cooperative agreement, and by the CARMA partner
universities. Some of the data presented herein were obtained at the
W.M. Keck Observatory, which is operated as a scientific partnership
among the California Institute of Technology, the University of
California and the National Aeronautics and Space Administration. The
Observatory was made possible by the generous financial support of the
W.M. Keck Foundation. The authors wish to recognize and acknowledge
the very significant cultural role and reverence that the summit of
Mauna Kea has always had within the indigenous Hawaiian community. We
are most fortunate to have the opportunity to conduct observations
from this mountain.

{\it Facilities:} \facility{Keck:II}, \facility{VLT:Yepun},
\facility{IRAM:Interferometer}, \facility{CARMA}, \facility{CSO}.


\clearpage

\figcaption[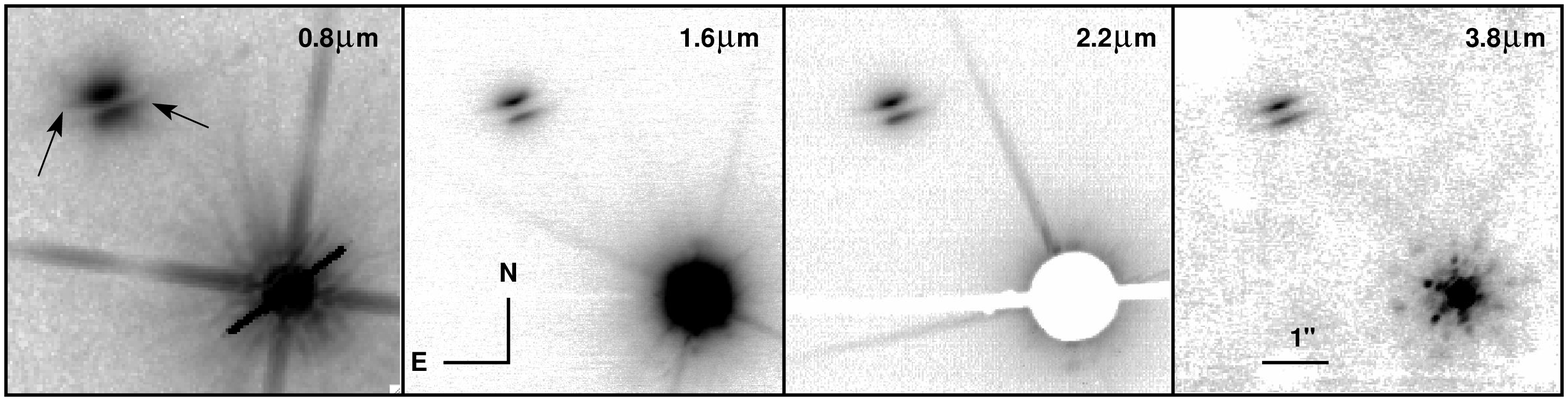]{Scattered light images of {\hv}: the left most
  panel shows the F814W image obtained by \citet{stapelfeldt03} while
  the other three panels show our new images obtained at $H$ and
  $K_{\rm s}$ with VLT/NACO (the $K_{\rm s}$ image is the
  coronagraphic one), and at $L'$ with Keck/NIRC2 and the LGS adaptive
  optics system. The F814W image is shown on the logarithmic stretch
  to better highlight low-level features. The arrows point to the
  ``rays'' identified by \citet{stapelfeldt03}. All other images are
  shown on a square root stretch. In all cases, the field-of-view is
  6\arcsec\ across. As discussed in the text, the $L'$ image has been
  smoothed and resampled to improve the signal-to-noise ratio without
  degrading its spatial resolution.
  \label{fig:newimgs}}

\figcaption[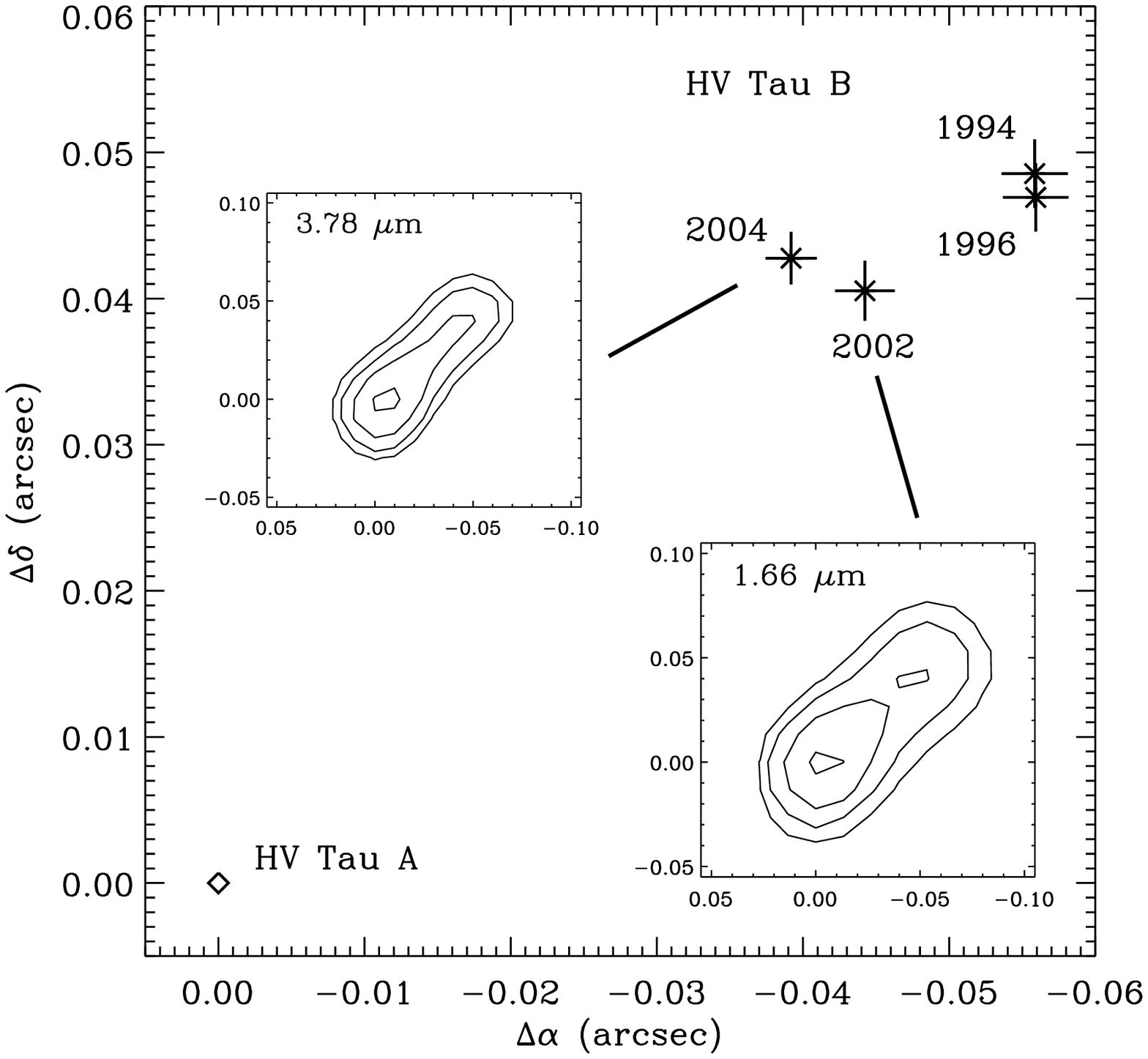]{Orbital motion of {\hvb} with respect to
  {\hva}, whose position is indicated by the diamond. The 1994 and
  1996 points reflect astrometric measurements by \citet{simon96} and
  \citet{monin00}, whereas the other two points are our 2002 and 2004
  measurements (see text). The insets represent our $H$ and $L'$
  images after a light Lucy deconvolution (25 and 50 iterations,
  respectively). While deconvolution better highlights the fact that
  the tight binary is resolved in our data, relative astrometry and
  photometry information for all our images was extracted by
  PSF-fitting of the original images. \label{fig:orbit}}

\figcaption[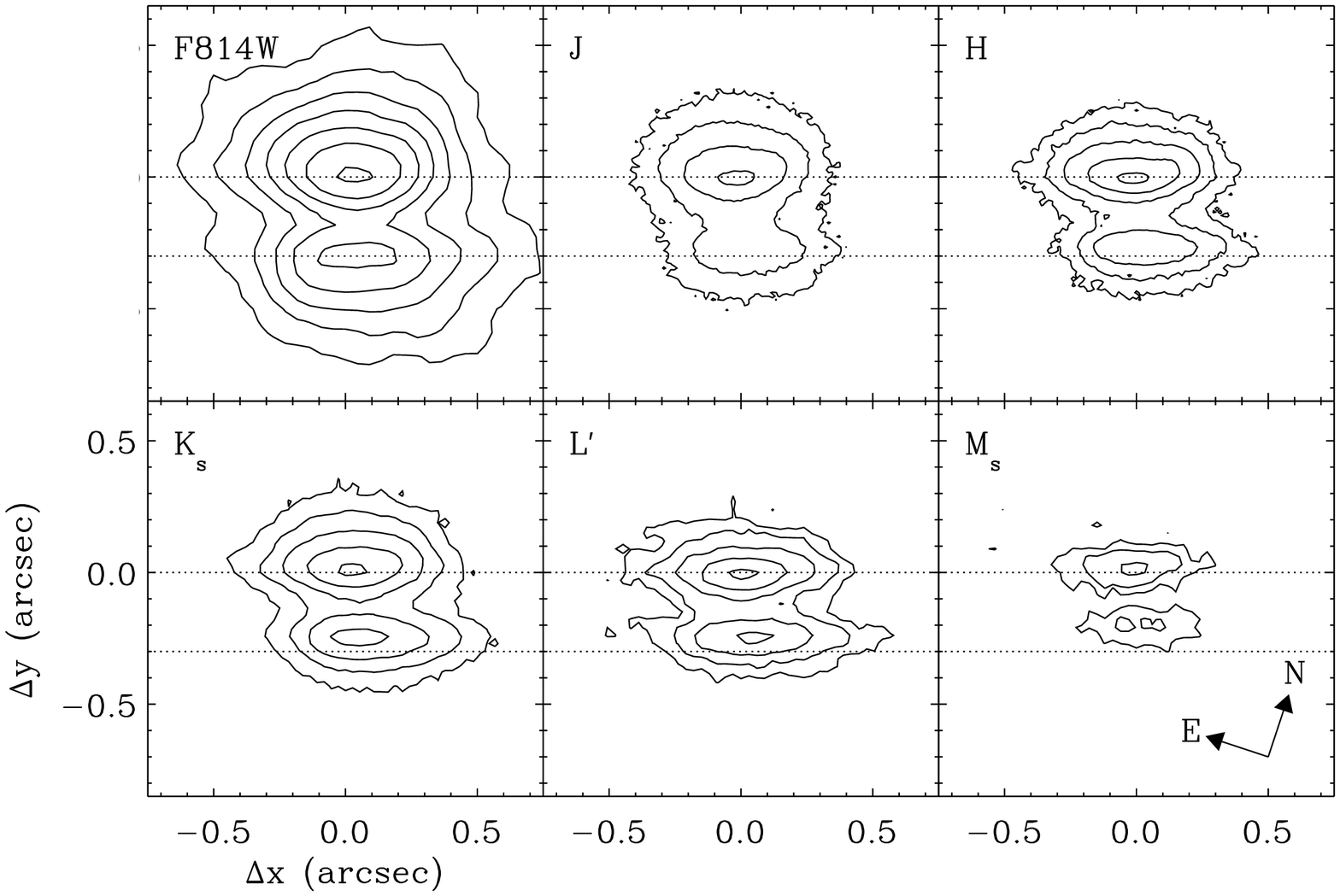]{Contour plots of {\hvc} after rotating all frames
  to a common orientation where the disk midplane is horizontal. The
  $K_{\rm s}$ image is the VLT/NACO coronagraphic one and the $L'$
  image is the Keck/NIRC2 LGS image. The contours lie from 90\% of the
  peak and by decreasing factors of 2 from there. The $L'$ and $M_{\rm
    s}$ images have been smoothed to improve their sensitivity (see
  Section 2.1.2 for more details). The dashed lines are guidelines
  indicating the location of the peak of each nebula in the F814W
  image.\label{fig:contours}}

\figcaption[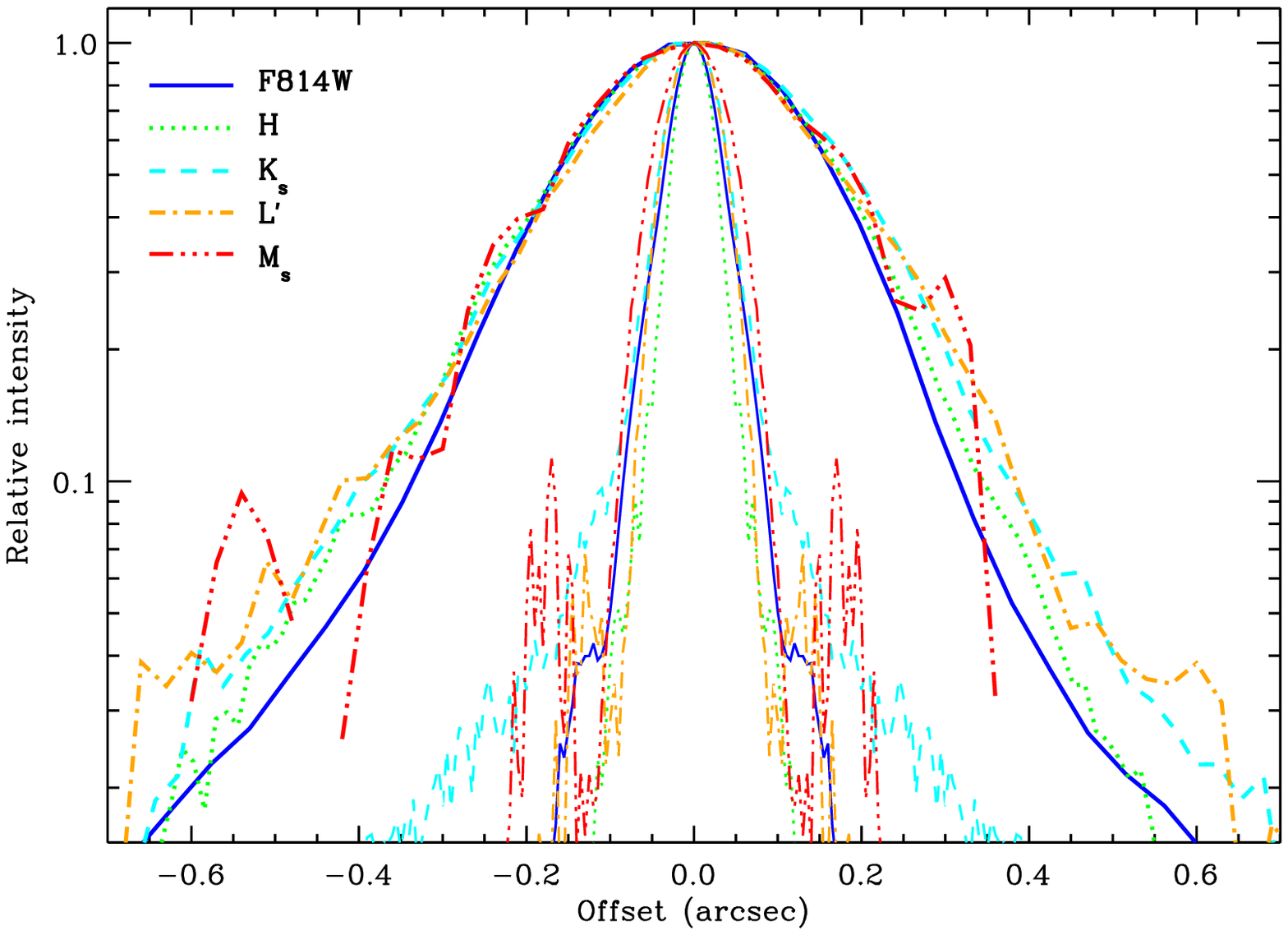]{Lateral intensity profiles along the brightest
  (NE) nebula of {\hvc}. Profiles are drawn for the F814W (solid
  blue), $H$ (dotted green), coronagraphic $K_{\rm s}$ (dashed cyan),
  LGS $L'$ (dot-dashed orange) and $M_{\rm s}$ (triple-dot-dashed red)
  images of the disk. All profiles are drawn normalizing the peak
  intensity in each image to unity. The thin curves represent the
  azimuthally-averaged profile of the PSF corresponding to each
  observation, using the same linestyles and color coding. The nebula
  is well resolved and its intensity profile is remarkably constant
  across the wavelength range probed here.\label{fig:profiles}}

\figcaption[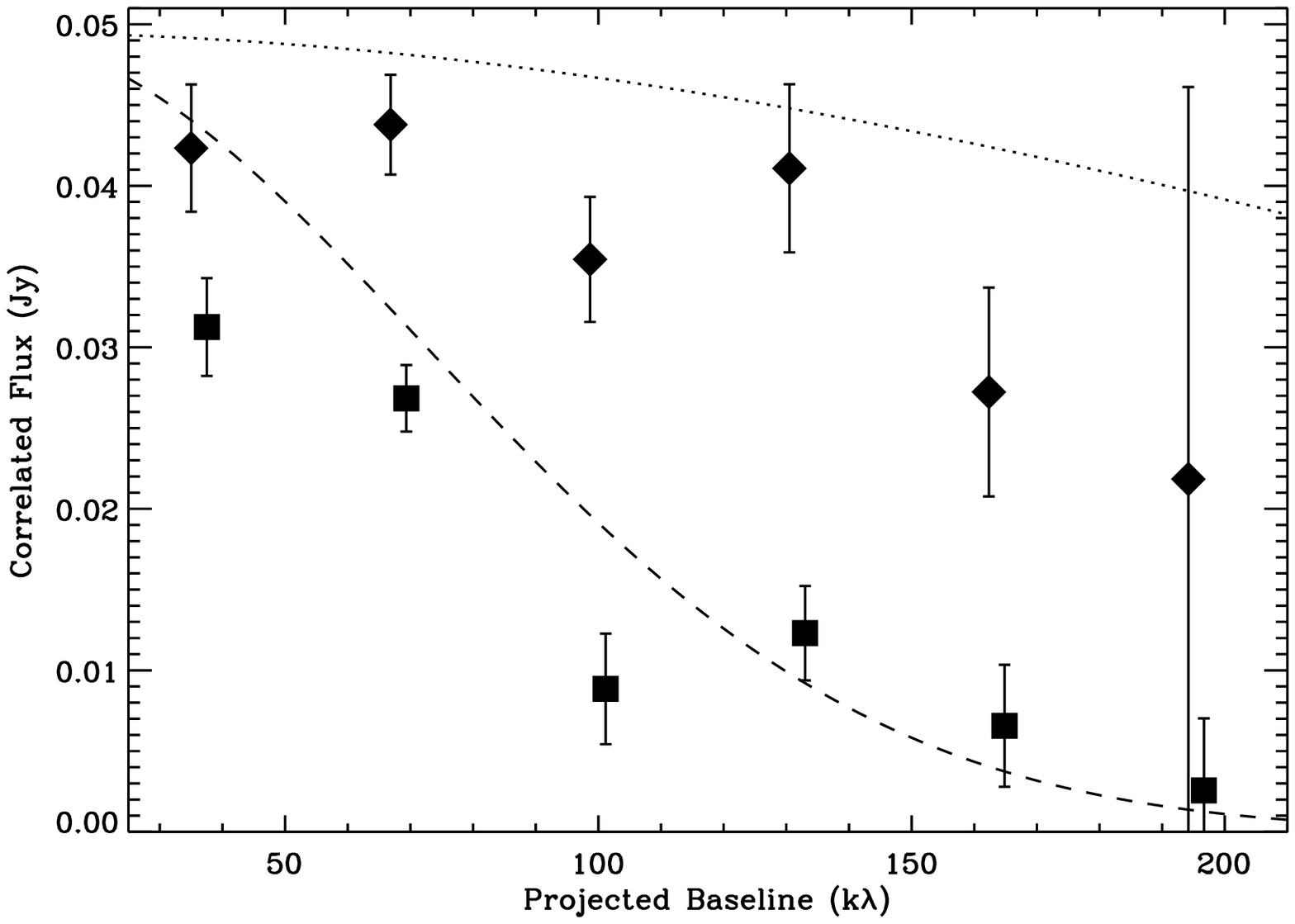]{PdBI 1.3\,mm correlated fluxes for {\hvc} as a
  function of projected baseline length for all baselines that are
  within 22\fdg5 of the disk major axis (squares) and within 22\fdg5
  of its minor axis (diamonds). The major axis is assumed to be
  108\fdg3 as determined from the scattered light images. The dashed
  and dotted curves represent the major- and minor-axis of the
  Gaussian fit to all visibilities, respectively, assuming a Gaussian
  0\farcs3 atmospheric ``seeing''. \label{fig:fcor}}

\figcaption[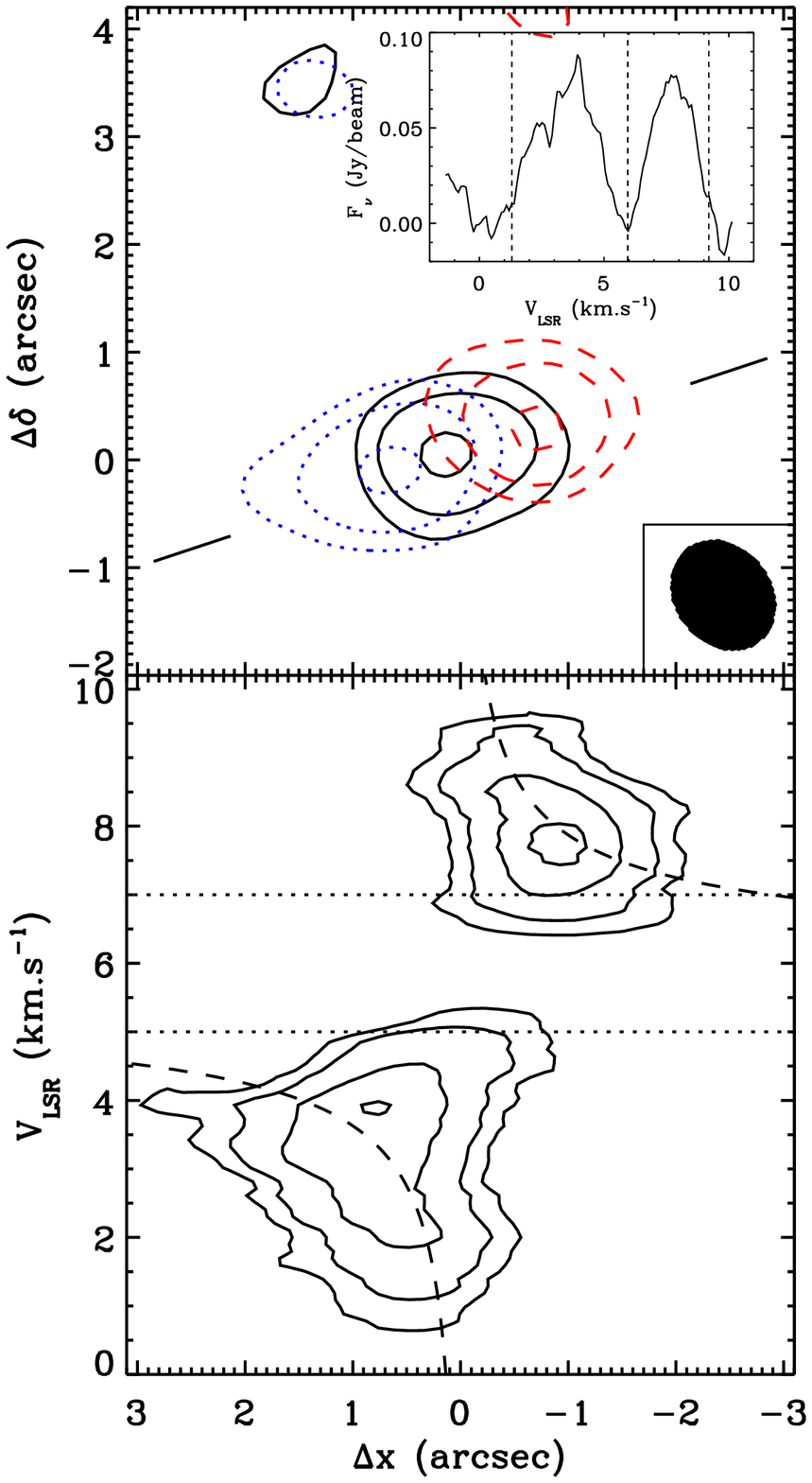]{{\it Top:} Contour plots of the blue (dotted
  contours, 0.8--6\,km.s$^{-1}$ velocity range) and red (dashed
  contours, 6--9.5\,km.s$^{-1}$ velocity range) parts of the $^{12}$CO
  (2--1) line, superimposed on the contours of the adjacent 1.3\,mm
  continuum (solid contours). Contours are drawn at 22.5, 45 and 90\%
  of the peak intensity in each map. The reconstructed beam is shown
  for comparison. The origin of the relative coordinates is at
  04h38m35.51s, +26\degr10\arcmin41\farcs5 (J2000), the nominal
  position of {\hvc}. The inset represents the integrated line
  profile. {\it Bottom:} Position-velocity diagram along the disk
  major axis (indicated by the two solid segments in the top
  panel). The three-dimensional datacube has been smoothed using a
  1\,km.s$^{-1}\times$1\arcsec\ running boxcar function. The contours
  in the position-velocity diagram are at 15, 30, 60 and 90\% of the
  peak intensity. The dashed curve represent the theoretical Keplerian
  rotation curve for an $0.7\,M_\odot$ star with $v_{\rm
    sys}=5.75$\,km.s$^{-1}$. This is not meant as a fit but a
  reference to guide the eye. The dotted lines indicate the velocity
  range in which $^{13}$CO emission from the Taurus molecular cloud
  was observed by \citet{mizuno95} at the location of
  {\hv}. \label{fig:co}}

\figcaption[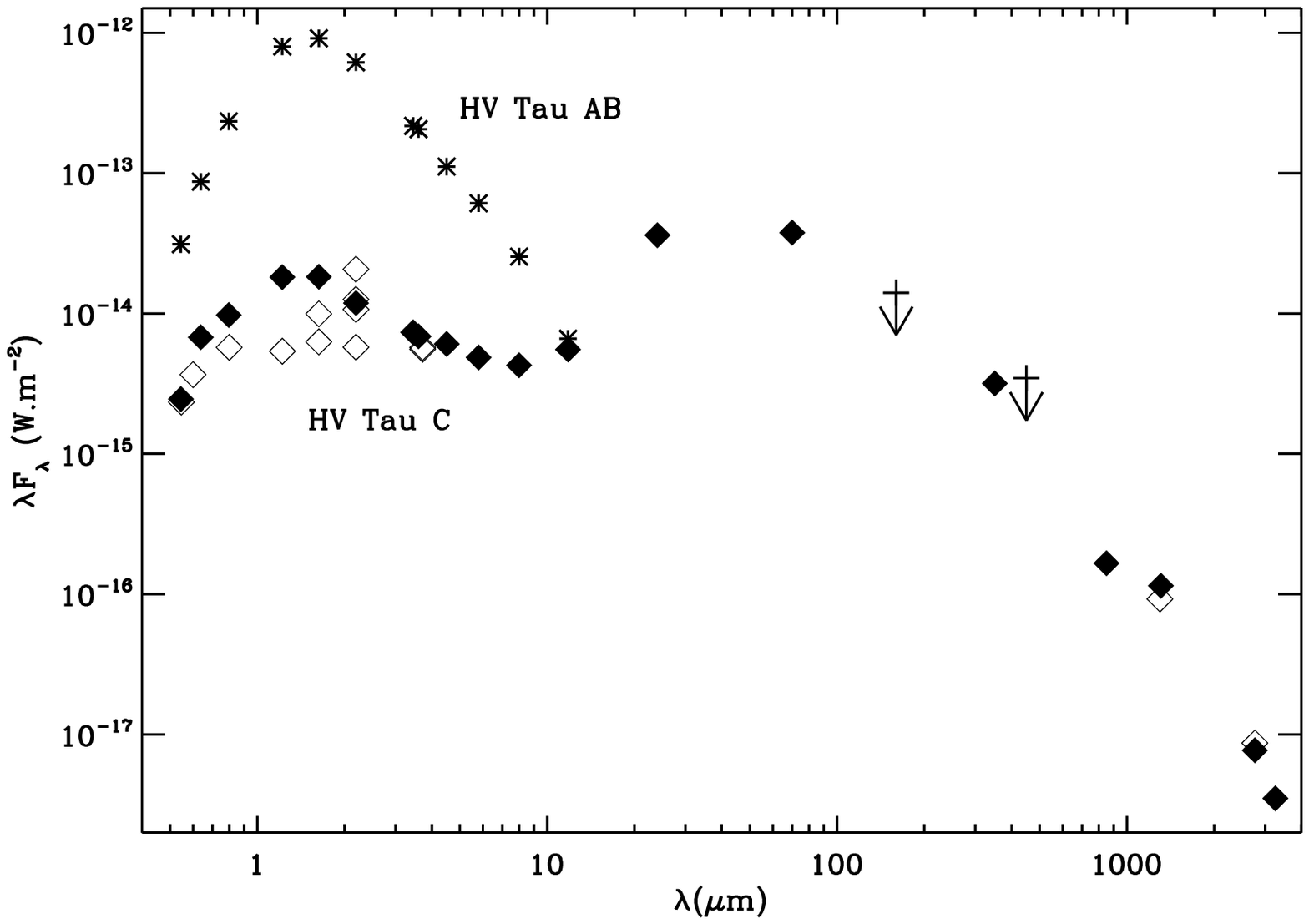]{SED of {\hvab} (asterisks) and {\hvc}
  (diamonds). Filled diamonds and upper limits indicate the photometry
  dataset considered in our model fitting (see Table~\ref{tab:photom})
  whereas empty diamonds represent photometry at other epochs,
  illustrating in particular the optical and near-infrared variability
  of {\hvc}. \label{fig:sed}}

\figcaption[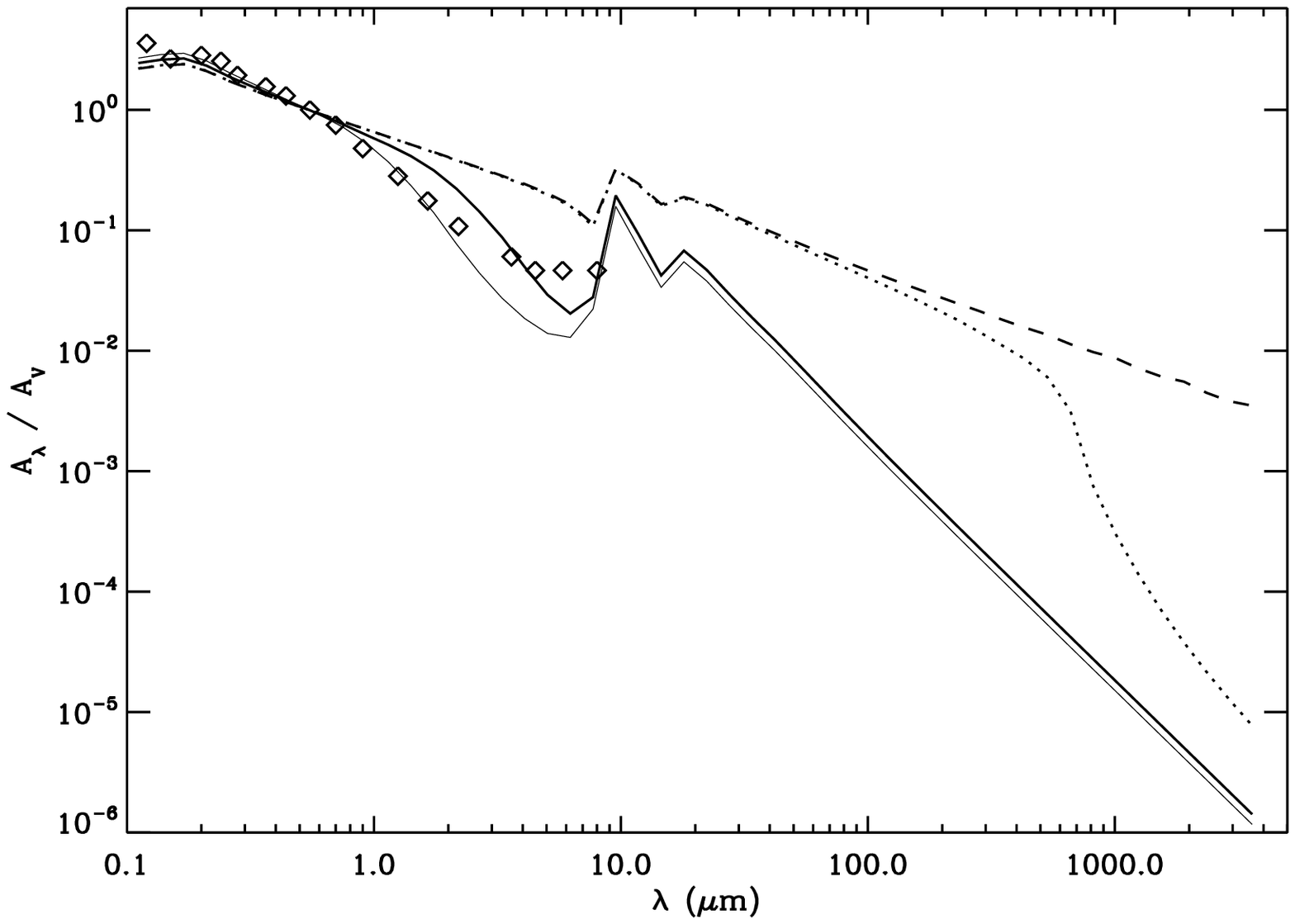]{Extinction law for our dust models assuming
  $a_{max}=1$\mic\ (thick solid), $a_{max}=100$\mic\ (thick dotted),
  and $a_{max}=1$cm (thick dashed). Diamonds represent the measured
  extinction law from \citet{ccm89} and \citet{indebetouw05}. The thin
  solid curve represents a model with $a_{max}=0.5$\mic (not used in
  our modeling) that best reproduces the measured interstellar
  extinction law up to 2\mic, but falls short of it in the 5--8\mic\
  regime. The $a_{max}=1$\mic\ is better in that regime and we use it
  as a proxy for ``interstellar dust'' in our
  model.\label{fig:opacity}}

\figcaption[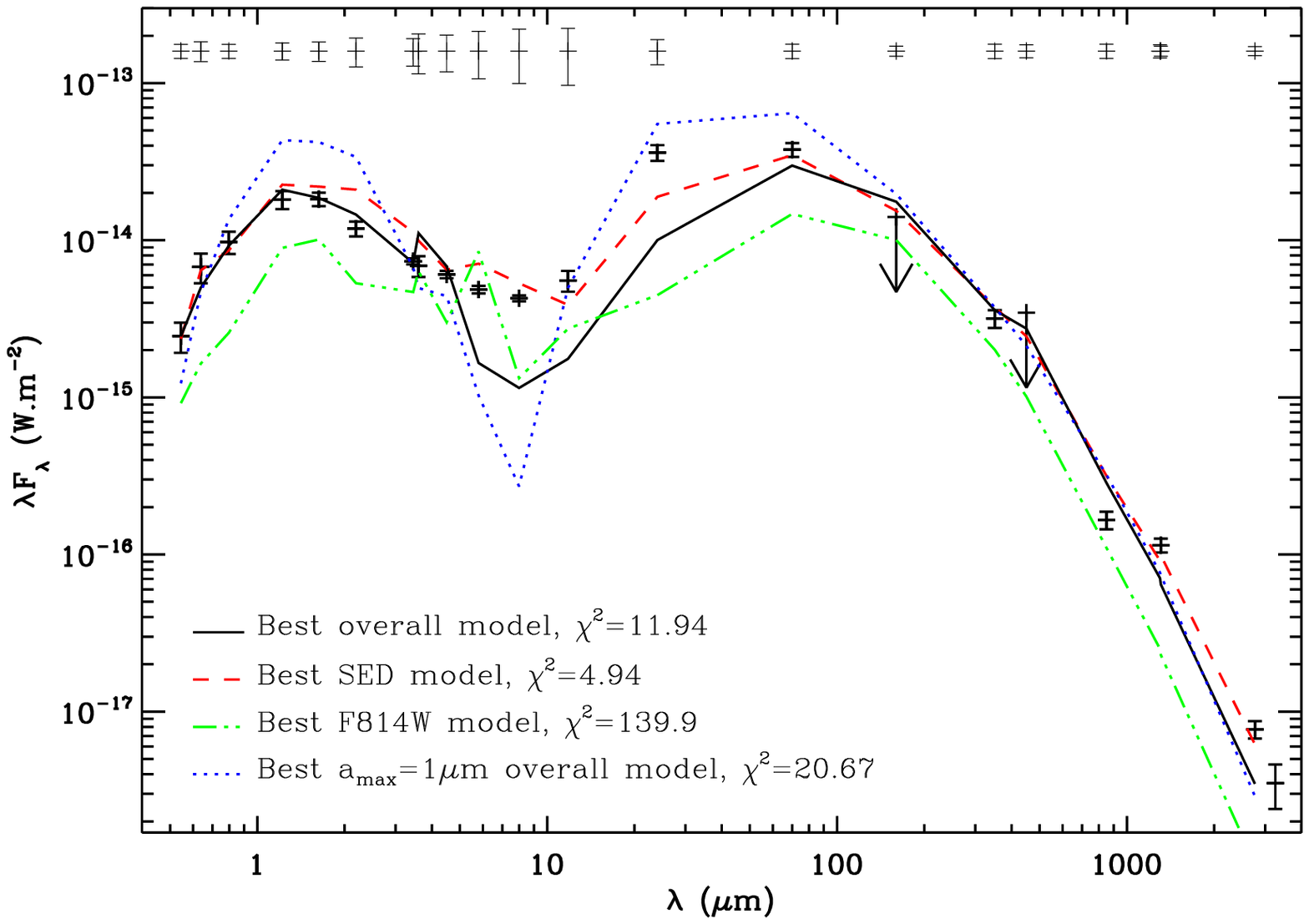]{Comparison of the SED of {\hvc} (from
  Table\,\ref{tab:photom}, see also Figure\,\ref{fig:sed}) with
  various models: the best fit to all available datasets ({\it black
    solid curve}), the best fit to the SED alone ({\it red dashed
    curve}), the best fit to the F814W image ({\it green dot-dashed
    curve}), and the best overall fit with $a_{\rm max}=1$\mic,
  i.e. interstellar-like dust properties ({\it blue dotted
    curve}). The thin errorbars indicated at the top of the plot
  represent our Monte Carlo noise estimate. This noise is apparent in
  the jumps seen at 3\mic\ for the best overall model and at 6\mic\
  for the best F814W fit. \label{fig:fit_sed}}

\figcaption[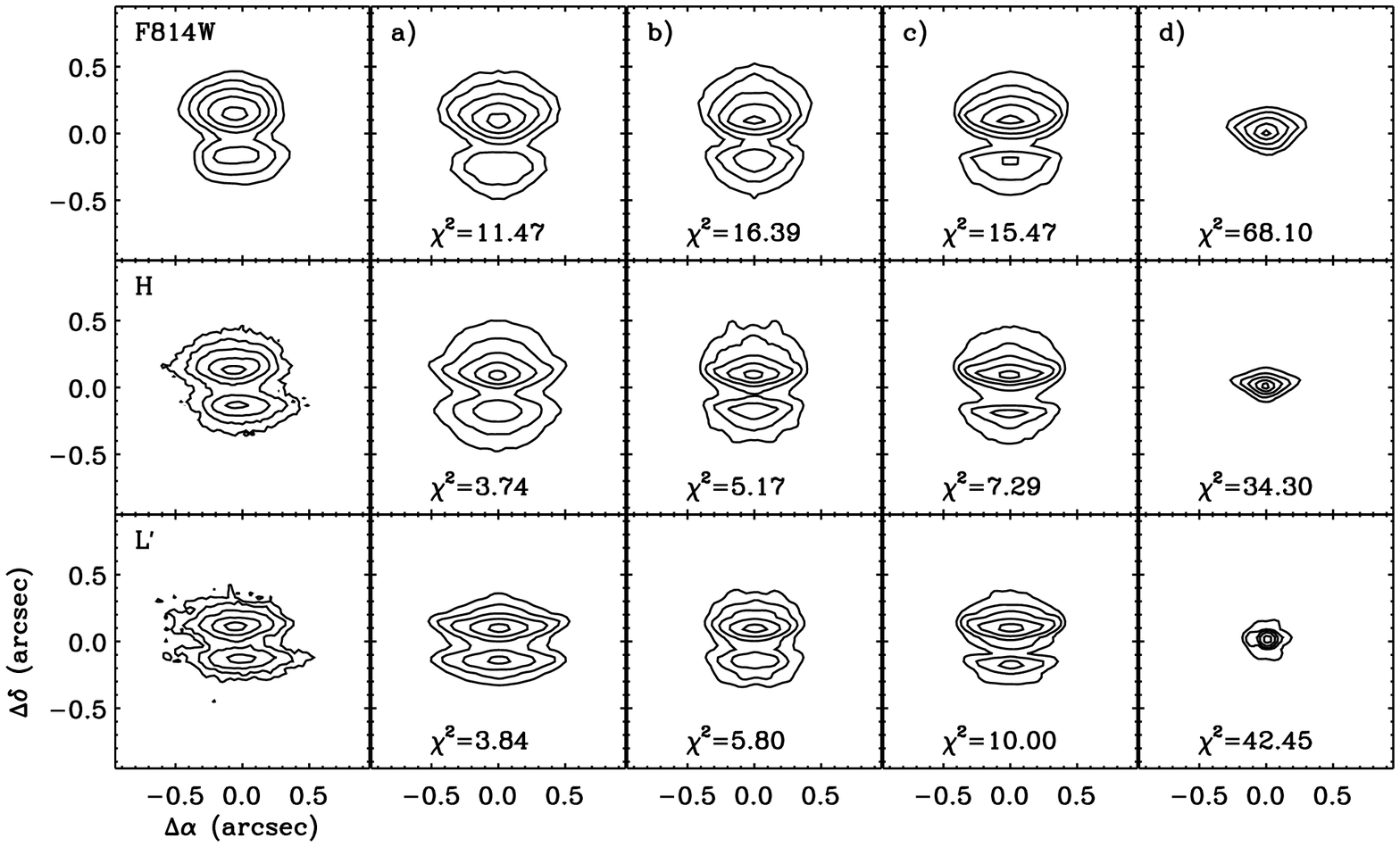]{Contour plots of {\hvc} at 0.8\mic\ ({\it top
    row}), 1.6\mic\ ({\it middle row}) and 3.8\mic\ ({\it bottom
    row}). From left to right, the columns represent: the actual data,
  the best model in our grid at each wavelength ({\it a}), the best
  model to all three scattered light images simultaneously ({\it b}),
  the best overall model ({\it c}), and the best fit to the SED alone
  ({\it d}). All model images have been convolved with the appropriate
  PSF. In each plot, the contours lie at 80, 40, 20, 10 and 5\% of the
  peak. \label{fig:fit_imgs}}

\figcaption[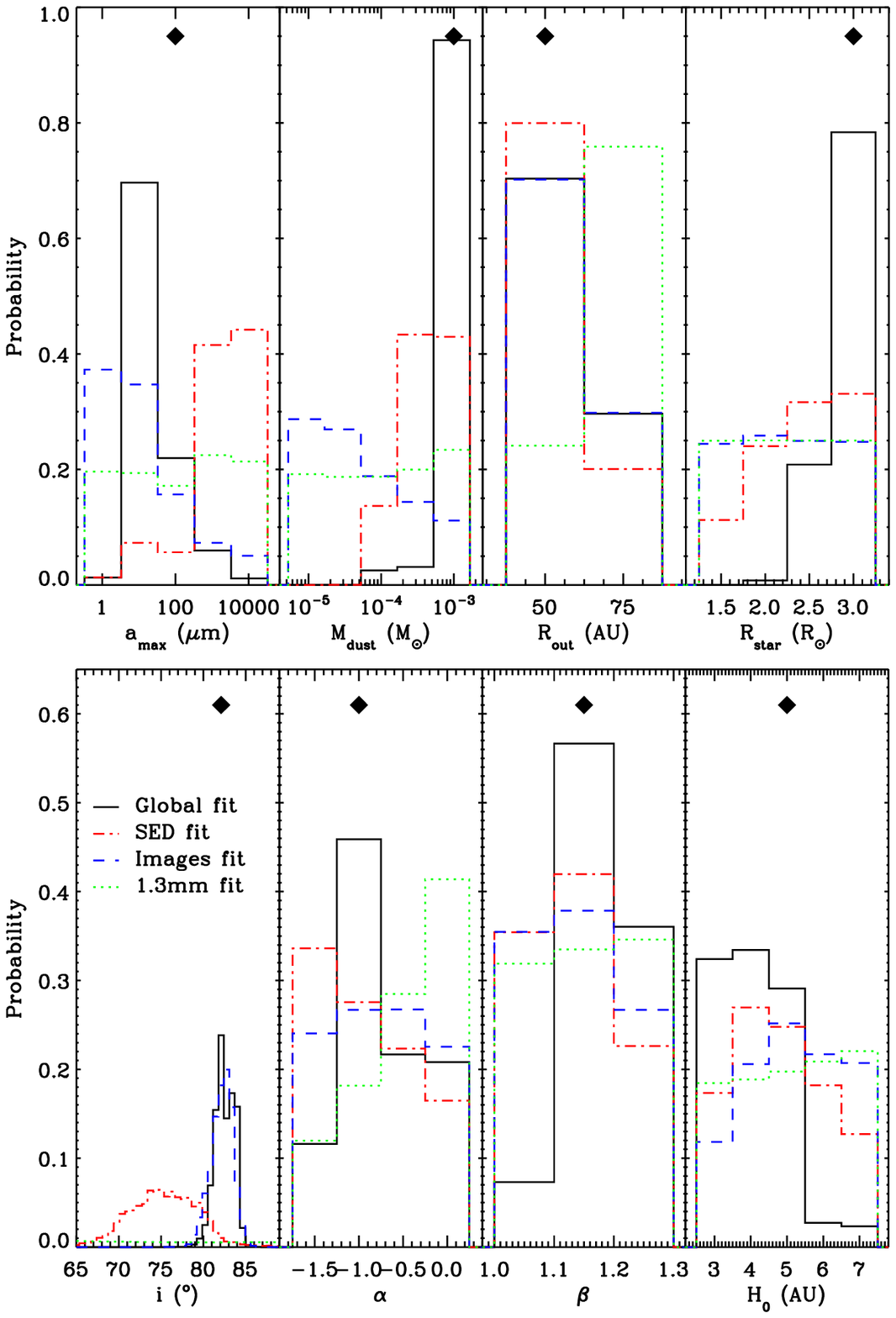]{Bayesian inference probability distributions for
  the free parameters in our model, based on fitting the three
  scattered light images at once ({\it dashed blue histograms}), the
  SED ({\it dot-dashed red}), the spatially resolved 1.3\,mm
  correlated fluxes ({\it dotted green}), and all observations at once
  ({\it solid black}). The probability distribution for the
  inclination based on the 1.3\,mm is almost flat and very close to
  0. For reference, the filled diamonds indicate the value of the
  parameters for the model with the absolute lowest $\chi^2_{\rm tot}$
  in our grid. Notice the different vertical scale in the two rows of
  plots. \label{fig:bayes_all}}

\clearpage

%
%

\begin{figure*}
  \plotone{f1.eps}
\end{figure*}

\begin{figure}
  \plotone{f2.eps}
\end{figure}

\begin{figure}
  \plotone{f3.eps}
\end{figure}

\begin{figure}
  \plotone{f4.eps}
\end{figure}

\begin{figure}
  \plotone{f5.eps}
\end{figure}

\begin{figure}
  \epsscale{0.8}
  \plotone{f6.eps}
\end{figure}

\begin{figure}
  \epsscale{1.0}
  \plotone{f7.eps}
\end{figure}

\begin{figure}
  \plotone{f8.eps}
\end{figure}

\begin{figure}
  \plotone{f9.eps}
\end{figure}

\begin{figure}
  \plotone{f10.eps}
\end{figure}

\begin{figure}
  \plotone{f11.eps}
\end{figure}

\clearpage

\begin{table}
\caption{Astrometric\tablenotemark{a} and photometric properties of {\hv}\label{tab:binary}}
\begin{tabular}{ccccccc}
  \hline
  $\lambda$ & Epoch & $\rho_{\rm AB-C}$ & $PA_{\rm AB-C}$ & $\rho_{\rm A-B}$ &
  $PA_{\rm A-B}$ & $\Delta m_{\rm A-B}$ \\
  (\mic) & & (\arcsec) & (\degr) & (\arcsec) & (\degr) & (mag) \\
  \hline
0.80 & 2000.19 & 4.04$\pm$0.01 & 43.4$\pm$0.3 & -- & -- & -- \\
1.27 & 2002.90 & 4.08$\pm$0.03 & 43.5$\pm$1.0 & -- & -- & -- \\
1.66 & 2002.90 & 4.06$\pm$0.03 & 43.6$\pm$1.0 & 0.054$\pm$0.003 & 309$\pm$3 &
0.88$\pm$0.02 \\ 
2.18 & 2002.90 & 4.03$\pm$0.03 & 43.6$\pm$1.0 & 0.062$\pm$0.004 & 317$\pm$2 &
0.88$\pm$0.05 \\ 
3.78 & 2002.95 & 4.03$\pm$0.01 & 44.9$\pm$0.1 & 0.063$\pm$0.003 & 312$\pm$2 &
0.53$\pm$0.04 \\ 
3.78 & 2004.84 & 4.02$\pm$0.01 & 44.4$\pm$0.1 & 0.058$\pm$0.002& 317.5$\pm$1.5 &
0.59$\pm$0.05 \\ 
4.67 & 2002.95 & 4.07$\pm$0.02 & 44.7$\pm$0.2 & -- & -- & -- \\
\hline
\end{tabular}
\tablenotetext{a}{Astrometric uncertainties include both measurement (including
  centroiding) and absolute calibration uncertainties.}
\end{table}

\begin{table}
\caption{Observed properties of the {\hvc} disk\label{tab:observables}}
\begin{tabular}{ccccccl}
  \hline
  $\lambda$ & $PA_{\rm disk}$ & $d_{\rm neb}$ & $FR_{\rm peak}$ &
  $FR_{\rm int}$ & $w_{\rm 5\sigma}$ & Note \\
  (\mic) & (\degr) & (\arcsec) & & & (\arcsec) & \\
  \hline
  0.80 & 108.2$\pm$1.2 & 0.335$\pm$0.005 &
  4.0$\pm$0.2 & 2.8$\pm$0.1 & 1.55$\pm$0.03 & \\ 
  1.27 & 107.5$\pm$1.1 & 0.281$\pm$0.002 &
  2.5$\pm$0.2 & 2.4$\pm$0.1 & 1.04$\pm$0.07 & \\ 
  1.66 & 108.4$\pm$0.9 & 0.284$\pm$0.002 &
  2.3$\pm$0.2 & 2.2$\pm$0.1 & 0.96$\pm$0.05 & \\ 
  2.18 & 108.4$\pm$0.7 & 0.275$\pm$0.003 &
  1.8$\pm$0.1 & 1.80$\pm$0.05 & 1.21$\pm$0.07 & direct image \\ 
  2.18 & 107.7$\pm$0.8 & 0.267$\pm$0.003 &
  2.0$\pm$0.1 & 1.82$\pm$0.05 & 1.24$\pm$0.07 & coronagraphic image\\ 
  3.78 & 108.8$\pm$1.3 & 0.237$\pm$0.004 &
  1.9$\pm$0.2 & 1.8$\pm$0.1 & 0.66$\pm$0.03 & NGS image \\
  3.78 & 109.1$\pm$0.9 & 0.236$\pm$0.003 &
  2.1$\pm$0.2 & 1.5$\pm$0.1 & 0.87$\pm$0.05 & LGS image \\
  4.67 & 108.2$\pm$1.7 & 0.231$\pm$0.006 &
  1.8$\pm$0.3 & 2.3$\pm$0.2 & 0.45$\pm$0.05 & \\
  \hline
\end{tabular}
\end{table}

\begin{table}
\caption{Adopted SED for {\hvc}\label{tab:photom}}
\begin{tabular}{ccccl|ccccl}
\hline
$\lambda$ & $F_\nu$ & $\sigma(F_\nu)$ & Date & Ref. & $\lambda$ & $F_\nu$ &
$\sigma(F_\nu)$ & Date & Ref. \\ 
 (\mic) & (mJy) & (mJy) & & & ($\mu$m) & (mJy) & (mJy) & \\
\hline
0.545 & 0.446 & 0.097 & 1991 Jan-Feb & 1 & 11.8 & 21.8 & 3.3 & 2002 Nov 13 & 3 \\
0.638 & 1.44 & 0.31 & 1991 Jan-Feb & 1 & 24. & 289. & 19. & 2005 Feb 28 & 5 \\
0.797 & 2.59 & 0.42 & 1991 Jan-Feb & 1 & 70. & 880 & 63 & 2005 Feb 28 & 5 \\
1.22 & 7.38 & 0.97 & 1996 Aug 26 & 2 & 160. & $<$750. & -- & 2005 Feb 28 & 5 \\
1.65 & 9.92 & 0.97 & 1996 Aug 26 & 2 & 350. & 370. & 30. & 2008 Jan 28 & 6 \\
2.19 & 8.66 & 0.94 & 2001 Dec 1 & 3 & 450. & $<$519. & -- & 2000 Feb 2 & 7 \\
3.45 & 8.44 & 0.37 & 2001 Dec 1 & 3 & 850. & 47. & 6. & 2000 Feb 2 & 7 \\
3.6 & 8.25 & 1.25 & 2004 Mar 7 & 4 & 1300. & 40. & 6. & 1988 Apr 29--30 & 8 \\
4.5 & 9.09 & 0.50 & 2004 Mar 7 & 4 & 2763 & 7.1 & 0.9 & 2005 Feb 26 & 6 \\
5.8 & 9.39 & 0.51 & 2004 Mar 7 & 4 & 3252 & 3.8 & 1.2 & 2008 May 29 & 6 \\
8. & 11.4 & 0.5 & 2004 Mar 7& 4 & & & & & \\
\hline
\end{tabular}
\tablerefs{1 -- \citet{magazzu94}; 2 -- \citet{woitas98}; 3 --
  \citet{mccabe06}; 4 -- \citet{hartmann05}; 5 -- {\it Spitzer} Taurus
  Legacy Survey (Rebull et al., in prep.); 6 -- this work; 7 --
  \citet{andrews05} ; 8 -- \citet{beckwith90}}
\end{table}

\begin{table}
\caption{Model parameters\label{tab:models}}
\begin{tabular}{lcccl}
  \hline
  Parameter & Min. value & Max. value & $N_{\rm sampl}$ & Sampling \\
  \hline
  \multicolumn{5}{c}{{\it Radiative transfer parameters}}\\
  \hline
  $M_{\rm dust}\,(M_\odot)$ & $10^{-5}$ & $10^{-3}$ & 5 & logarithmic \\
  $R_{\rm in}$~(AU) & \multicolumn{2}{c}{0.15} & -- & fixed \\
  $R_{\rm out}$~(AU) & 50 & 75 & 2 & linear \\
  $H_0$\tablenotemark{a}~(AU) & 3. & 7. & 5 & linear \\
  $\beta$ & 1.05 & 1.25 & 3 & linear \\
  $\alpha$ & -1.5 & 0. & 4 & linear \\
  $a_{\rm min}$~(\mic) & \multicolumn{2}{c}{0.03} & -- & fixed \\
  $a_{\rm max}$~(\mic) & 1. & $10^4$ & 5 & logarithmic \\
  $p$ & \multicolumn{2}{c}{3.7} & -- & fixed \\
  $R_\star~(R_\odot)$ & 1.5 & 3. & 4 & linear \\
  $T_\star$~(K) & \multicolumn{2}{c}{3800} & -- & fixed \\
  $i~(\degr)$ & 0. & 90. & 91 & linear in $\cos i$ \\
  \hline
  \multicolumn{5}{c}{{\it Post-processing parameters}}\\
  \hline
  $D$~(pc) & \multicolumn{2}{c}{140} & -- & fixed \\
  $A_V$~(mag) & 0 & 5. & 21 & linear \\
  \hline
\end{tabular}
\tablenotetext{a}{$H_0$ is defined at a radius $r_0=50$\,AU from the
  central star.} 
\end{table}

\begin{table}
  \caption{Goodness-of-fit estimates for selected models\label{tab:chi2}} 
\begin{tabular}{lccccccc}
\hline
& $\chi^2_{\rm F814W}$ & $\chi^2_{\rm H}$ & $\chi^2_{\rm L'}$ &
$\chi^2_{0.8-3.8\mu\mathrm{m}}$ & $\chi^2_{\rm SED}$ & $\chi^2_{\rm 1.3mm}$ &
$\chi^2_{\rm tot}$ \\ 
\hline
  Best F814W model & {\bf 11.47} & 9.24 & 17.69 & 38.40 & 139.89 &
  15.30 & 193.59 \\  
  Best $H$ model & 22.32 & {\bf 3.74} & 11.22 & 37.28 & 204.58 & 16.62
  & 258.47 \\ 
  Best $L'$ model & 36.64 & 5.97 & {\bf 3.84} & 46.45 & 126.97 & 14.53
  & 187.96 \\
  Best images model & 16.39 & 5.17 & 5.80 & {\bf 27.36} & 174.95 &
  15.87 & 218.17 \\
  Best SED model & 68.10 & 34.30 & 42.45 & 144.85 & {\bf 4.94} & 16.10
  & 165.89 \\
  Best 1.3\,mm model & 122.60 & 58.80 & 58.46 & 239.86 & 297.23 & {\bf
    9.19} & 546.28 \\  
  Best overall model & 15.47 & 7.29 & 10.00 & 32.76 & 11.94 & 16.35 &
  {\bf 61.04} \\
\hline
\end{tabular}
\end{table}


\begin{thebibliography}{}
\bibitem[Alves de Oliveira \& Casali(2008)]{alves08} Alves de
  Oliveira, C., \& Casali, M.\ 2008, \aap, 485, 155
\bibitem[Akeson et al.(2002)]{akeson02} Akeson, R.~L., Ciardi, D.~R.,
  van Belle, G.~T., \& Creech-Eakman, M.~J.\ 2002, \apj, 566, 1124
\bibitem[Andrews \& Williams (2005)]{andrews05} Andrews, S.~M., \&
  Williams, J.~P.\ 2005, \apj, 631, 1134 
\bibitem[Andrews \& Williams(2007)]{andrews07} Andrews, S.~M., \&
  Williams, J.~P.\ 2007, \apj, 659, 705
\bibitem[Appenzeller et al.(2005)]{appenzeller05} Appenzeller, I.,
  Bertout, C., \& Stahl, O.\ 2005, \aap, 434, 1005
\bibitem[Baraffe et al.(1998)]{baraffe98} Baraffe, I., Chabrier, G.,
  Allard, F., \& Hauschildt, P.~H.\ 1998, \aap, 337, 403
\bibitem[Barri{\`e}re-Fouchet et al.(2005)]{barriere05}
  Barri{\`e}re-Fouchet, L., Gonzalez, J.-F., Murray, J.~R., Humble,
  R.~J., \& Maddison, S.~T.\ 2005, \aap, 443, 185
\bibitem[Beckwith et al.(1990)]{beckwith90} Beckwith, S.~V.~W.,
  Sargent, A.~I., Chini, R.~S., \& Guesten, R.\ 1990, \aj, 99, 924
\bibitem[Beckwith \& Sargent(1991)]{beckwith91} Beckwith, S.~V.~W., \&
  Sargent, A.~I.\ 1991, \apj, 381, 250
\bibitem[Bertout et al.(1988)]{bertout88} Bertout, C., Basri, G., \&
  Bouvier, J.\ 1988, \apj, 330, 350
\bibitem[Bertout \& Genova(2006)]{bertout06} Bertout, C., \& Genova,
  F.\ 2006, \aap, 460, 499
\bibitem[Burrows et al.(1996)]{burrows96} Burrows, C.~J., et al.\
  1996, \apj, 473, 437
\bibitem[Cardelli et al.(1989)]{ccm89} Cardelli, J.~A., Clayton,
  G.~C., \& Mathis, J.~S.\ 1989, \apj, 345, 245
\bibitem[Chiang et al.(2001)]{chiang01} Chiang, E.~I., Joung, M.~K.,
  Creech-Eakman, M.~J., Qi, C., Kessler, J.~E., Blake, G.~A., \& van
  Dishoeck, E.~F.\ 2001, \apj, 547, 1077
\bibitem[Cotera et al.(2001)]{cotera01} Cotera, A.~S., et al.\ 2001,
  \apj, 556, 958
\bibitem[D'Alessio et al.(2001)]{dalessio01} D'Alessio, P., Calvet,
  N., \& Hartmann, L.\ 2001, \apj, 553, 321
\bibitem[D'Alessio et al.(2006)]{dalessio06} D'Alessio, P., Calvet,
  N., Hartmann, L., Franco-Hern{\'a}ndez, R., \& Serv{\'{\i}}n, H.\
  2006, \apj, 638, 314
\bibitem[Dowell et al.(2003)]{dowell03} Dowell, C.~D., et al.\ 2003,
  \procspie, 4855, 73
\bibitem[Draine (2003)]{draine03} Draine, B.~T.\ 2003, \apj, 598, 
1017 
\bibitem[Duch{\^e}ne et al.(2003)]{duchene03} Duch{\^e}ne, G.,
  M{\'e}nard, F., Stapelfeldt, K., \& Duvert, G.\ 2003, \aap, 400, 559
\bibitem[Duch{\^e}ne et al.(2004)]{duchene04} Duch{\^e}ne, G., McCabe,
  C., Ghez, A.~M., \& Macintosh, B.~A.\ 2004, \apj, 606, 969
\bibitem[Duch{\^e}ne et al.(2007)]{duchene07} Duch{\^e}ne, G.,
  Bontemps, S., Bouvier, J., Andr{\'e}, P., Djupvik, A.~A., \& Ghez,
  A.~M.\ 2007, \aap, 476, 229
\bibitem[Ducourant et al.(2005)]{ducourant05} Ducourant, C., Teixeira,
  R., P{\'e}ri{\'e}, J.~P., Lecampion, J.~F., Guibert, J., \& Sartori,
  M.~J.\ 2005, \aap, 438, 769
\bibitem[Dullemond \& Dominik(2004)]{dullemond04} Dullemond, C.~P., \&
  Dominik, C.\ 2004, \aap, 421, 1075
\bibitem[Dullemond \& Dominik(2005)]{dullemond05} Dullemond, C.~P., \&
  Dominik, C.\ 2005, \aap, 434, 971
\bibitem[Dutrey et al.(1996)]{dutrey96} Dutrey, A., Guilloteau, S.,
  Duvert, G., Prato, L., Simon, M., Schuster, K., \& Menard, F.\ 1996,
  \aap, 309, 493
\bibitem[Eggenberger et al.(2007)]{eggenberger07} Eggenberger, A.,
  Udry, S., Chauvin, G., Beuzit, J.-L., Lagrange, A.-M.,
  S{\'e}gransan, D., \& Mayor, M.\ 2007, \aap, 474, 273
\bibitem[Eiroa et al.(2002)]{eiroa02} Eiroa, C., et al.\ 2002, \aap,
  384, 1038
\bibitem[Flaherty et al.(2007)]{flaherty07} Flaherty, K.~M., Pipher,
  J.~L., Megeath, S.~T., Winston, E.~M., Gutermuth, R.~A., Muzerolle,
  J., Allen, L.~E., \& Fazio, G.~G.\ 2007, \apj, 663, 1069
\bibitem[Furlan et al.(2006)]{furlan06} Furlan, E., et al.\ 2006,
  \apjs, 165, 568
\bibitem[Ghez et al.(1995)]{ghez95} Ghez, A.~M., Weinberger, A.~J.,
  Neugebauer, G., Matthews, K., \& McCarthy, D.~W., Jr.\ 1995, \aj,
  110, 753
\bibitem[Ghez et al.(2008)]{ghez08} Ghez, A.~M., et al.\ 2008, \apj,
  689, 1044
\bibitem[Glauser et al.(2008)]{glauser08} Glauser, A.~M., M{\'e}nard,
  F., Pinte, C., Duch{\^e}ne, G., G{\"u}del, M., Monin, J.-L., \&
  Padgett, D.~L.\ 2008, \aap, 485, 531
\bibitem[Guilloteau et al.(1992)]{guilloteau92} Guilloteau, S., et
  al.\ 1992, \aap, 262, 624
\bibitem[Hartmann et al.(2005)]{hartmann05} Hartmann, L., Megeath,
  S.~T., Allen, L., Luhman, K., Calvet, N., D'Alessio, P.,
  Franco-Hernandez, R., \& Fazio, G.\ 2005, \apj, 629, 881
\bibitem[Hayashi(1981)]{hayashi81} Hayashi, C.\ 1981, Progress of
  Theoretical Physics Supplement, 70, 35
\bibitem[Hughes et al.(2008)]{hughes08} Hughes, A.~M., Wilner, D.~J.,
  Qi, C., \& Hogerheijde, M.~R.\ 2008, \apj, 678, 1119
\bibitem[Indebetouw et al.(2005)]{indebetouw05} Indebetouw, R., et
  al.\ 2005, \apj, 619, 931
\bibitem[Isella et al.(2007)]{isella07} Isella, A., Testi, L., Natta,
  A., Neri, R., Wilner, D., \& Qi, C.\ 2007, \aap, 469, 213
\bibitem[Keene \& Masson(1990)]{keene90} Keene, J., \& Masson, C.~R.\
  1990, \apj, 355, 635
\bibitem[Kenyon \& Hartmann(1987)]{kenyon87} Kenyon, S.~J., \&
  Hartmann, L.\ 1987, \apj, 323, 714
\bibitem[Kenyon \& Hartmann(1995)]{kenyon95} Kenyon, S.~J., \&
  Hartmann, L.\ 1995, \apjs, 101, 117
\bibitem[Kessler-Silacci et al.(2006)]{kessler06} Kessler-Silacci, J.,
  et al.\ 2006, \apj, 639, 275
\bibitem[Kim et al.(1994)]{kim94} Kim, S.-H., Martin, P.~G., \&
  Hendry, P.~D.\ 1994, \apj, 422, 164
\bibitem[Kimura et al.(2003)]{kimura03} Kimura, H., Kolokolova, L., \&
  Mann, I.\ 2003, \aap, 407, L5
\bibitem[Kitamura et al.(2002)]{kitamura02} Kitamura, Y., Momose, M.,
  Yokogawa, S., Kawabe, R., Tamura, M., \& Ida, S.\ 2002, \apj, 581,
  357
\bibitem[Kuchner(2004)]{kuchner04} Kuchner, M.~J.\ 2004, \apj, 612,
  1147
\bibitem[Laibe et al.(2008)]{laibe08} Laibe, G., Gonzalez, J.-F.,
  Fouchet, L., \& Maddison, S.~T.\ 2008, \aap, 487, 265
\bibitem[Lay et al.(1994)]{lay94} Lay, O.~P., Carlstrom, J.~E., Hills,
  R.~E., \& Phillips, T.~G.\ 1994, \apjl, 434, L75
\bibitem[Lay et al.(1997)]{lay97} Lay, O.~P., Carlstrom, J.~E., \&
  Hills, R.~E.\ 1997, \apj, 489, 917
\bibitem[Lenzen et al.(2003)]{lenzen03} Lenzen, R., et al.\ 2003,
  \procspie, 4841, 944
\bibitem[Magazzu \& Martin(1994)]{magazzu94} Magazzu, A., \& Martin,
  E.~L.\ 1994, \aap, 287, 571
\bibitem[Mannings \& Emerson(1994)]{mannings94} Mannings, V., \&
  Emerson, J.~P.\ 1994, \mnras, 267, 361
\bibitem[McCabe et al.(2003)]{mccabe03} McCabe, C., Duch{\^e}ne, G.,
  \& Ghez, A.~M.\ 2003, \apjl, 588, L113
\bibitem[McCabe et al.(2006)]{mccabe06} McCabe, C., Ghez, A.~M.,
  Prato, L., Duch{\^e}ne, G., Fisher, R.~S., \& Telesco, C.\ 2006,
  \apj, 636, 932
\bibitem[Mizuno et al.(1995)]{mizuno95} Mizuno, A., Onishi, T.,
  Yonekura, Y., Nagahama, T., Ogawa, H., \& Fukui, Y.\ 1995, \apjl,
  445, L161
\bibitem[Monin \& Bouvier(2000)]{monin00} Monin, J.-L., \& Bouvier,
  J.\ 2000, \aap, 356, L75
\bibitem[Mugrauer \& Neuh{\"a}user(2009)]{mugrauer09} Mugrauer, M., \&
  Neuh{\"a}user, R.\ 2009, \aap, 494, 373
\bibitem[Mundy et al.(1996)]{mundy96} Mundy, L.~G., et al.\ 1996,
  \apjl, 464, L169
\bibitem[Muzerolle et al.(2003)]{muzerolle03} Muzerolle, J., Calvet,
  N., Hartmann, L., \& D'Alessio, P.\ 2003, \apjl, 597, L149
\bibitem[Natta et al.(2000)]{natta00} Natta, A., Grinin, V., \&
  Mannings, V.\ 2000, Protostars and Planets IV, 559
\bibitem[Natta et al.(2007)]{natta07} Natta, A., Testi, L., Calvet,
  N., Henning, T., Waters, R., \& Wilner, D.\ 2007, Protostars and
  Planets V, 767
\bibitem[Pani\'c et al.(2009)]{panic09} Pani\'c, O., Hogerheijde,
  M.~R., Wilner, D., \& Qi, C.\ 2009, \aap, in press
  (astro-ph/0904.1127)
\bibitem[Pi{\'e}tu et al.(2005)]{pietu05} Pi{\'e}tu, V., Guilloteau,
  S., \& Dutrey, A.\ 2005, \aap, 443, 945
\bibitem[Pinte et al.(2006)]{pinte06} Pinte, C., M{\'e}nard, F.,
  Duch{\^e}ne, G., \& Bastien, P.\ 2006, \aap, 459, 797
\bibitem[Pinte et al.(2007)]{pinte07} Pinte, C., Fouchet, L.,
  M{\'e}nard, F., Gonzalez, J.-F., \& Duch{\^e}ne, G.\ 2007, \aap,
  469, 963
\bibitem[Pinte et al.(2008)]{pinte08} Pinte, C., et al.\ 2008,
A\&A, 489, 633
\bibitem[Pontoppidan et al.(2007)]{pontoppidan07} Pontoppidan, K.~M.,
  Stapelfeldt, K.~R., Blake, G.~A., van Dishoeck, E.~F., \& Dullemond,
  C.~P.\ 2007, \apjl, 658, L111
\bibitem[Roddier et al.(1996)]{roddier96} Roddier, C., Roddier, F.,
  Northcott, M.~J., Graves, J.~E., \& Jim, K.\ 1996, \apj, 463, 326
\bibitem[Rodmann et al.(2006)]{rodmann06} Rodmann, J., Henning, T.,
  Chandler, C.~J., Mundy, L.~G., \& Wilner, D.~J.\ 2006, \aap, 446,
  211
\bibitem[Rousset et al.(2003)]{rousset03} Rousset, G., et al.\ 2003,
  \procspie, 4839, 140
\bibitem[Sauter et al.(2009)]{sauter09} Sauter, J., et al.\ 2009,
  \aap, in press
\bibitem[Simon et al.(1992)]{simon92} Simon, M., Chen, W.~P., Howell,
  R.~R., Benson, J.~A., \& Slowik, D.\ 1992, \apj, 384, 212
\bibitem[Simon et al.(1996)]{simon96} Simon, M., Holfeltz, S.~T., \&
  Taff, L.~G.\ 1996, \apj, 469, 890
\bibitem[Stapelfeldt et al.(1998)]{stapelfeldt98} Stapelfeldt, K.~R.,
  Krist, J.~E., Menard, F., Bouvier, J., Padgett, D.~L., \& Burrows,
  C.~J.\ 1998, \apjl, 502, L65
\bibitem[Stapelfeldt \& Moneti(1999)]{stapelfeldt99} Stapelfeldt, K.,
  \& Moneti, A.\ 1999, The Universe as Seen by ISO, 427, 521
\bibitem[Stapelfeldt et al.(2003)]{stapelfeldt03} Stapelfeldt, K.~R.,
  M{\'e}nard, F., Watson, A.~M., Krist, J.~E., Dougados, C., Padgett,
  D.~L., \& Brandner, W.\ 2003, \apj, 589, 410
\bibitem[Strom et al.(1989)]{strom89} Strom, K.~M., Newton, G.,
  Strom, S.~E., Seaman, R.~L., Carrasco, L., Cruz-Gonzalez, I.,
  Serrano, A., \& Grasdalen, G.~L.\ 1989, \apjs, 71, 183
\bibitem[Strom \& Strom(1994)]{strom94} Strom, K.~M., \& Strom, S.~E.\
  1994, \apj, 424, 237
\bibitem[Terada et al.(2007)]{terada07} Terada, H., Tokunaga, A.~T.,
  Kobayashi, N., Takato, N., Hayano, Y., \& Takami, H.\ 2007, \apj,
  667, 303
\bibitem[van Dam et al.(2006)]{vandam06} van Dam, M.~A., et al.\ 2006,
  \pasp, 118, 310
\bibitem[Watson \& Stapelfeldt(2004)]{watson04} Watson, A.~M., \&
  Stapelfeldt, K.~R.\ 2004, \apj, 602, 860
\bibitem[Watson \& Stapelfeldt(2007)]{watson07} Watson, A.~M., \&
  Stapelfeldt, K.~R.\ 2007, \aj, 133, 845
\bibitem[Watson et al.(2007)]{watson07ppv} Watson, A.~M., Stapelfeldt,
  K.~R., Wood, K., \& M{\'e}nard, F.\ 2007, Protostars and Planets V,
  523
\bibitem[Weidenschilling(1997)]{weidenschilling97} Weidenschilling,
  S.~J.\ 1997, Icarus, 127, 290
\bibitem[Weingartner \& Draine(2001)]{weingartner01} Weingartner,
  J.~C., \& Draine, B.~T.\ 2001, \apj, 548, 296
\bibitem[White \& Ghez(2001)]{white01} White, R.~J., \& Ghez, A.~M.\
  2001, \apj, 556, 265
\bibitem[Wizinowich et al.(2000)]{wizi00} Wizinowich, P., et al.\
  2000, \pasp, 112, 315
\bibitem[Wizinowich et al.(2006)]{wizi06} Wizinowich, P.~L., et al.\
  2006, \pasp, 118, 297
\bibitem[Woitas \& Leinert(1998)]{woitas98} Woitas, J., \& Leinert, C.\
  1998, \aap, 338, 122
\bibitem[Wolf et al.(2003)]{wolf03} Wolf, S., Padgett, D.~L., \&
  Stapelfeldt, K.~R.\ 2003, \apj, 588, 373
\bibitem[Wood et al.(2002)]{wood02} Wood, K., Wolff, M.~J., Bjorkman,
  J.~E., \& Whitney, B.\ 2002, \apj, 564, 887
\bibitem[Wright(1987)]{wright87} Wright, E.~L.\ 1987, \apj, 320, 818
\bibitem[Zsom \& Dullemond(2008)]{zsom08} Zsom, A., \& Dullemond,
  C.~P.\ 2008, \aap, 489, 931
\end{thebibliography}
\end{document}